\documentclass[acmsmall,screen]{acmart} %DIF > 
\usepackage{booktabs} % For formal tables
\usepackage{makecell}
\usepackage[utf8]{inputenc}
\usepackage{array}
\usepackage{wrapfig}
\usepackage{multirow}
\usepackage{tabularx}
\usepackage{tabulary}
\newcolumntype{X}{L}

\usepackage{pifont}
\usepackage{lineno}
\usepackage{newunicodechar}
\newunicodechar{）}{)}
\usepackage{balance}
\usepackage{xcolor,pifont}
\definecolor{cyan}{HTML}{009aa8}

\newcommand{\revcolor}{black}
\newcommand{\rev}[1]{{\color{\revcolor}#1}}
\usepackage[linesnumbered, ruled,vlined]{algorithm2e}
\RequirePackage{pdflscape}
\usepackage{flushend}
\usepackage{multicol}
\usepackage{placeins}
\usepackage{indentfirst}
\usepackage{fancybox}
\usepackage[most]{tcolorbox}
\usepackage{mathtools}
\usepackage{rotating}
\usepackage{adjustbox}
\usepackage{colortbl}
\usepackage{color}
\usepackage{tcolorbox}
\usepackage{enumitem}
\usepackage{longtable}
\usepackage{ragged2e}
\newtcolorbox{mybox}[2][]
{colback = white, colframe = black, fonttitle = \bfseries,
    colbacktitle = gray, enhanced,
    attach boxed title to top left={yshift=-3mm, xshift=3mm},
    title=#2, #1}

\usepackage{tikz}
\usepackage{xcolor}

\DeclareRobustCommand{\circledzero}{%
  \tikz[baseline=(char.base)]%
    \node[circle, fill=black, inner sep=0.6pt] (char) {%
      \textcolor{white}{\scriptsize 0}%
    };%
}

\usepackage{framed}
\definecolor{Gray}{gray}{0.9}
\definecolor{shadecolor}{gray}{0.95}
\usetikzlibrary{trees,positioning,shapes,shadows,arrows}
\tikzset{
  basic/.style  = {draw, text width=2cm, drop shadow, font=\sffamily, rectangle},
  root/.style   = {basic, rounded corners=2pt, thin, align=center, fill=white},
  level-2/.style = {basic, rounded corners=6pt, thin,align=center, fill=white, text width=3cm},
  level-3/.style = {basic, thin, align=center, fill=white, text width=1.8cm}
}
\newcommand{\todo}[1]{}
\newcommand{\fakesection}[2][1em]{\vspace{#1}\noindent\textit{\textbf{#2}}}

\usepackage[normalem]{ulem}

\usepackage{amsmath}
\usepackage{pgfplots}
\pgfplotsset{compat=1.18}
\usepgfplotslibrary{groupplots}
\usepackage{subcaption}
\usepackage{tikz}
\usetikzlibrary{positioning}
\usepackage{algorithm2e}
\usepackage{xspace}
\usepackage{listings}
\usepackage{ragged2e}
\usepackage{makecell}
\newcommand{\tool}{\textsc{CodeTeam}\xspace}
\newcommand{\codes}{\textsc{CodeS}\xspace}
\newcommand{\sketcheval}{\textsc{SketchEval}\xspace}
\newcommand{\sketchbleu}{\textsc{SketchBLEU}\xspace}
\newcommand{\nlrepobench}{\textsc{NL2Repo-Bench}\xspace}

\newtcolorbox{rqbox}{
  colback=gray!7,
  colframe=gray!40,
  boxrule=0.5pt,
  arc=2pt,
  left=6pt,
  right=6pt,
  top=6pt,
  bottom=6pt
}

\newtcolorbox{rqsummary}[1][]{
  colback=gray!8,
  colframe=gray!50,
  boxrule=0.8pt,
  arc=6pt,
  left=10pt,
  right=10pt,
  top=8pt,
  bottom=8pt,
  before upper={\textbf{#1}: \itshape}
}

\newcommand{\agentrole}[1]{\textsc{#1}}
\providecommand{\PassAtOne}{Pass@1} %DIF PREAMBLE
\providecommand{\texorpdfstring}[2]{#1} %DIF PREAMBLE
\DeclareRobustCommand{\DIFadd}[1]{\texorpdfstring{\protect\leavevmode{\rev{#1}}}{#1}} %DIF PREAMBLE
\providecommand{\DIFdel}[1]{} %DIF PREAMBLE
\providecommand{\DIFaddbegin}{} %DIF PREAMBLE
\providecommand{\DIFaddend}{} %DIF PREAMBLE
\providecommand{\DIFaddFL}[1]{\DIFadd{#1}} %DIF PREAMBLE
\providecommand{\DIFdelFL}[1]{} %DIF PREAMBLE
\providecommand{\DIFaddbeginFL}{} %DIF PREAMBLE
\providecommand{\DIFaddendFL}{} %DIF PREAMBLE
\providecommand{\BluePunctUnskip}{\unskip} %DIF PREAMBLE
\DeclareRobustCommand{\BlueNewFloat}{\color{\revcolor}\ifdefined\captionsetup\captionsetup{labelfont={color=\revcolor},textfont={color=\revcolor}}\fi} %DIF PREAMBLE
\providecommand{\BlueBibBegin}{\color{\revcolor}} %DIF PREAMBLE
\providecommand{\BlueBibEnd}{\color{black}\ifdefined\hypersetup\hypersetup{citecolor=ACMPurple,linkcolor=ACMPurple,urlcolor=ACMDarkBlue,filecolor=ACMDarkBlue}\fi} %DIF PREAMBLE
\makeatletter %DIF PREAMBLE
\providecommand{\BlueBibSelectKey}[1]{\ifcsname BlueBibKey@#1\endcsname\BlueBibBegin\else\BlueBibEnd\fi} %DIF PREAMBLE
\def\BlueWrappedLbibitem[#1]#2{\BlueBibSelectKey{#2}\BlueOriginalLbibitem[#1]{#2}} %DIF PREAMBLE
\AtBeginDocument{\@ifpackageloaded{natbib}{\let\BlueOriginalLbibitem\@lbibitem\let\@lbibitem\BlueWrappedLbibitem\let\BlueOriginalEndBibliography\endthebibliography\def\endthebibliography{\BlueBibEnd\BlueOriginalEndBibliography}}{}} %DIF PREAMBLE
\makeatother %DIF PREAMBLE
\AtBeginDocument{\ifdefined\hypersetup\hypersetup{colorlinks=true,citecolor=ACMPurple,linkcolor=ACMPurple,urlcolor=ACMDarkBlue,filecolor=ACMDarkBlue}\fi} %DIF PREAMBLE
\expandafter\def\csname BlueBibKey@Dong2024SelfCollaboration\endcsname{1} %DIF PREAMBLE
\expandafter\def\csname BlueBibKey@Huang2023AgentCoder\endcsname{1} %DIF PREAMBLE
\expandafter\def\csname BlueBibKey@Qwen2026Qwen38\endcsname{1} %DIF PREAMBLE
\expandafter\def\csname BlueBibKey@Xia2025Agentless\endcsname{1} %DIF PREAMBLE
\expandafter\def\csname BlueBibKey@Zhang2024AutoCodeRover\endcsname{1} %DIF PREAMBLE
\makeatletter %DIF PREAMBLE
\let\sout@orig\sout %DIF PREAMBLE
\renewcommand{\sout}[1]{\ifmmode\text{\sout@orig{\ensuremath{#1}}}\else\sout@orig{#1}\fi} %DIF PREAMBLE
\makeatother %DIF PREAMBLE
\RequirePackage{listings} %DIF PREAMBLE
\lstdefinelanguage{DIFcode}{ %DIF PREAMBLE
  moredelim=[il][\color{white}\tiny]{\%DIF\ <\ }, %DIF PREAMBLE
  moredelim=[il][\sffamily\bfseries]{\%DIF\ >\ } %DIF PREAMBLE
} %DIF PREAMBLE
\lstdefinestyle{DIFverbatimstyle}{ %DIF PREAMBLE
	language=DIFcode, %DIF PREAMBLE
	basicstyle=\ttfamily, %DIF PREAMBLE
	columns=fullflexible, %DIF PREAMBLE
	keepspaces=true %DIF PREAMBLE
} %DIF PREAMBLE
\lstnewenvironment{DIFverbatim}[1][]{\lstset{style=DIFverbatimstyle,#1}}{} %DIF PREAMBLE
\lstnewenvironment{DIFverbatim*}[1][]{\lstset{style=DIFverbatimstyle,showspaces=true,#1}}{} %DIF PREAMBLE
\DeclareRobustCommand{\BlueNumber}[1]{\DIFadd{#1}}
\ifdefined\DeclareCaptionLabelFormat
\DeclareCaptionLabelFormat{bluenumber}{#1~\rev{#2}}
\fi
\providecommand{\BlueCaptionNumber}{\ifdefined\captionsetup\captionsetup{labelformat=bluenumber}\fi}
\makeatletter
\let\BlueOriginalSecCntFormat\@seccntformat
\providecommand{\BlueSectionNumberBegin}{\def\@seccntformat##1{\rev{\BlueOriginalSecCntFormat{##1}}}}
\providecommand{\BlueSectionNumberEnd}{\let\@seccntformat\BlueOriginalSecCntFormat}
\makeatother
\providecommand{\BlueBibNumberFormat}[1]{\ifcsname BlueBibNumber@#1\endcsname\rev{\BlueOriginalBibnumfmt{#1}}\else\BlueOriginalBibnumfmt{#1}\fi}
\AtBeginDocument{\ifdefined\bibnumfmt\let\BlueOriginalBibnumfmt\bibnumfmt\let\bibnumfmt\BlueBibNumberFormat\fi}
\expandafter\def\csname BlueBibNumber@6\endcsname{1}
\expandafter\def\csname BlueBibNumber@7\endcsname{1}
\expandafter\def\csname BlueBibNumber@9\endcsname{1}
\expandafter\def\csname BlueBibNumber@10\endcsname{1}
\expandafter\def\csname BlueBibNumber@11\endcsname{1}
\expandafter\def\csname BlueBibNumber@12\endcsname{1}
\expandafter\def\csname BlueBibNumber@13\endcsname{1}
\expandafter\def\csname BlueBibNumber@14\endcsname{1}
\expandafter\def\csname BlueBibNumber@15\endcsname{1}
\expandafter\def\csname BlueBibNumber@29\endcsname{1}
\expandafter\def\csname BlueBibNumber@30\endcsname{1}
\expandafter\def\csname BlueBibNumber@31\endcsname{1}
\expandafter\def\csname BlueBibNumber@32\endcsname{1}
\expandafter\def\csname BlueBibNumber@33\endcsname{1}
\expandafter\def\csname BlueBibNumber@34\endcsname{1}
\expandafter\def\csname BlueBibNumber@35\endcsname{1}
\expandafter\def\csname BlueBibNumber@37\endcsname{1}
\expandafter\def\csname BlueBibNumber@38\endcsname{1}
\expandafter\def\csname BlueBibNumber@39\endcsname{1}
\expandafter\def\csname BlueBibNumber@40\endcsname{1}
\expandafter\def\csname BlueBibNumber@41\endcsname{1}
\expandafter\def\csname BlueBibNumber@43\endcsname{1}
\expandafter\def\csname BlueBibNumber@44\endcsname{1}
\begin{document}
\title{CodeTeam: An LLM-Powered Multi-Agent Framework for Repository-Level Code Generation}

\author{Yifei Wang}
\orcid{0000-0003-0100-6896}
\email{whiten@whu.edu.cn}
\affiliation{
  \institution{School of Computer Science, Wuhan University}
  \city{Wuhan}
  \country{China}
}

\author{Ruiyin Li}
\orcid{0000-0001-8536-4935}
%\authornote{Corresponding author}
\affiliation{%
  \institution{School of Computer Science, Wuhan University}
  \city{Wuhan}
  \country{China}
}
 \DIFaddbegin \email{ryli\_cs@whu.edu.cn}
\DIFaddend 

\author{Peng Liang}
\orcid{0000-0002-2056-5346}
%\authornote{Corresponding author}
\affiliation{%
  \institution{School of Computer Science, Wuhan University}
  \city{Wuhan}
  \country{China}
}
\email{liangp@whu.edu.cn}

\author{Qiong Feng}
\orcid{0000-0003-1667-8062}
\affiliation{%
  \institution{School of Computer Science, Nanjing University of Science and Technology}
  \city{Nanjing}
  \country{China}
}
\email{qiongfeng@njust.edu.cn}

\author{Zengyang Li}
\orcid{0000-0002-7258-993X}
\affiliation{%
  \institution{School of Computer Science, Central China Normal University}
  \city{Wuhan}
  \country{China}
}
\email{zengyangli@ccnu.edu.cn}

\author{Mojtaba Shahin}
\orcid{0000-0002-9081-1354}
\affiliation{%
  \institution{School of Computing Technologies, RMIT University}
  \city{Melbourne}
  \country{Australia}
}
\email{mojtaba.shahin@rmit.edu.au}

\author{Arif Ali Khan}
\orcid{0000-0002-8479-1481}
\affiliation{%
  \institution{M3S Research Group, SEIS Unit, University of Oulu}
  \city{Oulu}
  \country{Finland}
}
\email{arif.khan@oulu.fi}

\acmJournal{TOSEM}
\acmVolume{0}
\acmNumber{0}
\acmArticle{0}
\acmMonth{0}

\renewcommand{\shortauthors}{Wang et al.}

\begin{abstract}
Natural language to repository generation (NL2Repo) requires a system to construct an entire software repository from a natural-language requirements document. Compared with function-level code generation, this task demands longer planning horizons, stable interfaces across files, and iterative debugging of cross-file inconsistencies. To address these challenges, we propose \tool, an LLM-based multi-agent framework that separates planning, decision making, and implementation into distinct, coordinated stages. In the planning stage, multiple \agentrole{Architect} agents draft competing software design sketches (SDSs), optionally grounded by retrieved design references. A \agentrole{cto} (Chief Technical Officer) agent then evaluates, selects, and normalizes the most promising SDS into a machine-checkable contract that specifies file ownership, public interfaces, and dependency constraints. In the implementation stage, \agentrole{Developer} agents generate code under a dependency-aware scheduler with bounded context and lightweight Git-based coordination, while a \agentrole{qa} agent runs tests and drives iterative repairs. On the synthesis-based \sketcheval benchmark, we \DIFaddbegin \DIFadd{compare }\DIFaddend \tool's prompt-engineering (PE) \DIFaddbegin \DIFadd{variant with }\DIFaddend \codes \DIFaddbegin \DIFadd{and five representative baselines across 19 repositories under a retrieval-free protocol, where }\DIFaddend \tool improves the \DIFaddbegin \DIFadd{mean structural score by 3.84 }\DIFaddend absolute points \DIFaddbegin \DIFadd{over \codes (53.88 versus 50.04)}\DIFaddend . On the execution-based \nlrepobench benchmark, used as an external validation protocol \DIFaddbegin \DIFadd{across 104 tasks}\DIFaddend , \tool achieves \DIFaddbegin \DIFadd{a mean counted-test pass rate of 48.24\% compared with 43.25\% for \codes, while full-task resolution remains 0/520 across all evaluated workflows and paired differences relative to the strongest adapter baseline span zero}\DIFaddend . Ablation results show that retrieval-augmented planning \DIFaddbegin \DIFadd{and SDS contract governance improve structural quality, while live interface briefs and \agentrole{qa}-driven repair contribute most to repository coherence and executable test passing}\DIFaddend . \tool and the experimental results are available at~\BlueNumber{\cite{replpack}}.
\end{abstract}

\ccsdesc[500]{Software and its engineering~Software development techniques} % [CCS Classification] https://dl.acm.org/ccs

\keywords{Large Language Model, Multi-Agent System, Repository-Level Code Generation, Software Engineering Automation, NL2Repo}

\maketitle

\section{Introduction}
Large Language Models (LLMs) have demonstrated strong capabilities in code generation, but prevailing evaluation paradigms remain largely confined to short, self-contained tasks such as single-function synthesis in HumanEval-style settings~\cite{Chen2021Codex}. In contrast, generating a complete software repository from natural-language requirements is a distinct task rather than a direct extension of function-level code generation, introducing challenges in file structures, modules, and interfaces~\BlueNumber{\cite{Zan2025CodeS}}. Recent execution-based \nlrepobench results further show that from-scratch repository generation requires architectural design, dependency management, multi-module implementation, and packaging, and that current agents exhibit failures such as loss of global architectural consistency and brittle cross-file dependencies~\BlueNumber{\cite{Ding2025NL2RepoBench}}. Consequently, generating a complete repository from a natural-language specification demands more than localized code completion. It requires a system that can design a complete repository structure, maintain  \DIFaddbegin \DIFadd{consistency }\DIFaddend among cross-file interfaces, and iteratively detect and repair errors that manifest throughout the subsequent lifecycle.

Recent repository-level benchmarks show that multi-file tasks are heterogeneous. RepoBench \cite{Liu2023RepoBench} and CrossCodeEval \BlueNumber{\cite{Ding2023CrossCodeEval}} focus on repository-level completion within an existing codebase. The FEA-Bench benchmark~\cite{Li2025FEABench} evaluates incremental feature implementation, requiring the model to add new functionality while editing related parts of an existing repository. Moreover, \codes~\BlueNumber{\cite{Zan2025CodeS}} and the more recent \nlrepobench \BlueNumber{\cite{Ding2025NL2RepoBench}} push one step further: the model starts from an empty workspace and must generate a complete repository solely from natural-language requirements. Across these settings, recurring challenges include managing long-range dependencies, reasoning under partial or underspecified information, and addressing errors that only manifest during system-level integration. These challenges indicate that repository-level code generation is not solely a modeling problem, but also a workflow design problem. Multi-agent software development systems, such as ChatDev~\cite{Qian2023ChatDev} and MetaGPT~\BlueNumber{\cite{Hong2023MetaGPT}}, demonstrate that role specialization and structured interaction can enhance multi-step generation. Similarly, repository-interaction agents such as CodeAgent~\BlueNumber{\cite{Zhang2024CodeAgent}}, SWE-agent~\BlueNumber{\cite{Yang2024SWEagent}}, and OpenHands~\BlueNumber{\cite{Wang2024OpenDevin}} further underscore the importance of tool use, environment interaction, and iterative execution for achieving reliable outcomes. 

Despite these advances, a critical \textbf{gap} remains in the natural language-to-repository (NL2Repo)~\BlueNumber{\cite{Zan2025CodeS}} generation. Existing approaches  address  \DIFaddbegin \DIFadd{complementary parts }\DIFaddend of the NL2Repo pipeline,  \DIFaddbegin \DIFadd{with different assumptions about planning, implementation, and repair}\DIFaddend . ChatDev~\cite{Qian2023ChatDev} and MetaGPT~\BlueNumber{\cite{Hong2023MetaGPT}} emphasize collaborative development workflows; SWE-agent~\BlueNumber{\cite{Yang2024SWEagent}} focuses on issue resolution in existing repositories; OpenHands~\BlueNumber{\cite{Wang2024OpenDevin}} provides an open platform for AI developers based on agents with integrated benchmarking; and CodeAgent~\BlueNumber{\cite{Zhang2024CodeAgent}} targets repo-level code generation with tool support for information retrieval, implementation, and testing. More closely related efforts, such as ProjectGen~\BlueNumber{\cite{Zhao2025ProjectGen}}, incorporate architecture design and skeleton construction. \DIFaddbegin \DIFadd{Building on these efforts, we focus on }\DIFaddend \textbf{three challenges} that arise when generating a repository from an empty workspace: (1)~high-level design planning that produces a machine-checkable contract, (2)~cross-file interface consistency maintained throughout implementation, and (3)~coordinating bounded-context implementation against that contract with dependency-aware scheduling and iterative repair.

To address this gap, we propose \tool, an LLM-based multi-agent framework tailored to NL2Repo. Our approach extends the sketch-based paradigm of \codes into a complete end-to-end workflow. Specifically, multiple \agentrole{Architect} agents first generate competing software design sketches (SDSs), optionally supported by retrieved design references. A dedicated \agentrole{cto} (Chief Technical Officer) agent subsequently selects the most promising design and normalizes it into a machine-checkable contract before code implementation. Then, \agentrole{Developer} agents implement files under a dependency-aware scheduler, \DIFaddbegin \DIFadd{and }\DIFaddend exchange interface updates through a lightweight Git-based workflow. Throughout this process, a \agentrole{qa} (Quality Assurance) agent is activated under certain conditions, such as when generation tasks are complete, or when test failures are detected. Detected failures are then fed back to the \agentrole{Developer} agents for repair.

The main \textbf{contributions} of this study are as follows: 
\begin{itemize}
    \item We propose \tool, a multi-agent framework for repo-level code generation that combines SDS competition, optional RAG (Retrieval-Augmented Generation) grounding, \agentrole{cto}-guided contract normalization, project-specific developer allocation, dependency-aware execution, Git-based coordination, and \agentrole{qa}-driven repair. 
    \item We conduct a comprehensive evaluation of \tool on \DIFaddbegin \DIFadd{the }\DIFaddend \sketcheval benchmark~\BlueNumber{\cite{Zan2025CodeS}}\DIFaddbegin \DIFadd{, }\DIFaddend comparing  \DIFaddbegin \DIFadd{its }\DIFaddend prompt-engineering (PE) variant  against  \codes  and  \DIFaddbegin \DIFadd{representative single-agent and multi-agent }\DIFaddend baselines. \DIFaddbegin \DIFadd{We report paired task-level comparisons and targeted ablations to examine the contributions of planning and implementation policies. 
    }\DIFaddend \item We introduce an external execution-based validation protocol on \nlrepobench to enable more rigorous end-to-end assessment.
\end{itemize}

\textbf{Organization}: The rest of this paper is structured as follows: Section~\ref{sec:related} presents the related work. Section~\ref{sec:study-design} describes the study design, including the overall framework of \tool, its core modules, and implementation details. Section~\ref{sec:results} presents the results, which are further analyzed and discussed in Section~\ref{sec:discussion}. The potential threats to validity are clarified in Section~\ref{sec:threats}. Finally, Section~\ref{sec:conclusion} concludes this work with future work directions.

\section{Related Work}\label{sec:related}

\subsection{From NL2Code to Long-Context Code Generation}
Previous work on natural language to code (NL2Code) mainly focuses on generating small, self-contained programs. Benchmarks such as CoNaLa~\BlueNumber{\cite{Yin2018Conala}}, CONCODE~\cite{Iyer2018Concode}, and APPS~\BlueNumber{\cite{Hendrycks2021APPS}} pair natural-language descriptions with target programs and evaluate NL2Code performance using metrics ranging from string-based similarity to execution-based tests. These datasets also drove the development of code-pretrained models. For example, CodeT5 introduces identifier-aware pre-training to improve both code understanding and code generation~\BlueNumber{\cite{Wang2021CodeT5}}, while the HumanEval framework established an influential test-based evaluation for synthesized programs~\cite{Chen2021Codex}. Open-source models then explore richer generation capabilities. InCoder explores bidirectional context and code infilling~\BlueNumber{\cite{Fried2022InCoder}}, whereas CodeGen emphasizes multi-turn program synthesis~\cite{Nijkamp2022CodeGen}.

Despite these advances, such settings largely underrepresent the complexity of real-world software development. In practice, software implementation activities must follow extensive surrounding context, including dependencies across files, shared utilities, build conventions, and repository-specific APIs. Although CONCODE can partially address this by providing the rest of the Java class as context~\cite{Iyer2018Concode}, most early NL2Code benchmarks still abstract away cross-file calls, configuration, and repository structure. Consequently, strong performance on isolated code snippets does not necessarily translate into the ability to generate coherent and consistent code at the repository level.

Recent work has begun to address this limitation by shifting attention from isolated generation to effective context utilization. Retrieval-augmented generation (RAG) introduces mechanisms for dynamically selecting relevant information as external memory~\cite{Lewis2020RAG}, and this idea has been adapted to code generation by retrieving relevant documentation at generation time~\BlueNumber{\cite{Zhou2023DocPrompting}}. Meanwhile, long-context adaptation methods such as LongLoRA extend usable context length at lower cost~\cite{Chen2023LongLoRA}. However, for repository-level code generation, increasing context length alone is insufficient. Effective systems must also determine which contextual information (e.g., design constraints and interface definitions) should be exposed at each step and ensure that such information remains consistent across multiple interdependent generation steps.

\subsection{Repository-Level Code Completion and NL2Repo}
Repository-level code generation can be divided into three related yet distinct settings, primarily differentiated by the extent to which the repository context is predefined versus constructed during generation. The first is repository-level completion, where the repository already exists and the main challenge is to retrieve and use relevant cross-file context. For example, RepoBench~\cite{Liu2023RepoBench} formalizes this setting and covers retrieval, completion, and end-to-end pipelines. CrossCodeEval~\BlueNumber{\cite{Ding2023CrossCodeEval}} broadens evaluations to multiple programming languages and shows that cross-file reasoning is beyond a single ecosystem. Methods such as RepoCoder~\BlueNumber{\cite{Zhang2023RepoCoder}} and GraphCoder~\cite{Liu2024GraphCoder} tailor retrieval to this setting by iteratively retrieving relevant repository snippets or code-context graphs before generation.

The second setting is repository-level incremental development, in which a system extends an existing repository by implementing a new feature. Compared to completion, this setting requires not only generating new code but also coordinating edits across multiple existing files. For example, FEA-Bench~\cite{Li2025FEABench} formalizes this feature-implementation setting and shows that current models struggle when required to add new components and modify existing ones under executable tests. Although this setting is closer to real-world maintenance scenarios than snippet generation, it still assumes that the repository structure, build configuration, and many interfaces are already in place.

The third setting is from-scratch repository generation from natural language requirements (NL2Repo). \codes~\BlueNumber{\cite{Zan2025CodeS}} formalizes this NL2Repo task and proposes a multi-layer sketch decomposition (RepoSketcher, FileSketcher, and SketchFiller), together with the \sketcheval benchmark and the \sketchbleu metric. More recently, \nlrepobench extends the from-scratch setting to 104 Python library tasks and evaluates generated repositories using the original upstream \texttt{pytest} suites, making the benchmark much stricter about end-to-end executability~\BlueNumber{\cite{Ding2025NL2RepoBench}}. 

The three settings above form a spectrum defined by the degree of available structure: from fully specified repositories (completion), to partially specified ones (incremental development), to entirely unconstrained generation (NL2Repo). While closely connected, they are not interchangeable. Methods that perform well when a repository structure and dependencies are given often fail when required to generate file trees, public APIs, and the dependency structure.

\subsection{LLM-Based Agents for Software Development}
A growing body of work investigates LLM-based agents that address complex tasks through multi-step interaction, tool use, and collaboration. In the context of software development, systems such as ChatDev~\cite{Qian2023ChatDev} organize specialized agents for design, coding, and testing, using structured communication protocols to improve end-to-end coherence. Similarly, MetaGPT~\BlueNumber{\cite{Hong2023MetaGPT}} incorporates standardized operating procedures into the prompting, reducing error propagation across development stages. More recent systems such as MapCoder~\cite{Islam2024MapCoder} further decompose multi-step code generation through explicit planning and retrieval-augmented agents. \DIFaddbegin \DIFadd{Along similar lines, Self-Collaboration~\BlueNumber{\cite{Dong2024SelfCollaboration}} organizes analyst, coder, and tester roles, while AgentCoder~\BlueNumber{\cite{Huang2023AgentCoder}} separates programming, test design, and test execution with iterative feedback. In repository issue resolution, AutoCodeRover~\BlueNumber{\cite{Zhang2024AutoCodeRover}} uses program structure to guide search and repair, whereas Agentless~\BlueNumber{\cite{Xia2025Agentless}} uses a simpler localization, repair, and validation pipeline; these approaches motivate explicit accounting for workflow complexity and cost, although their existing-repository setting differs from NL2Repo. }\DIFaddend Therefore, these systems show that role specialization and explicit coordination can improve the stability of long-horizon generation over single-pass generation tasks.

Recent studies evaluate LLM agents in realistic software engineering environments grounded in existing repositories. SWE-bench~\cite{Jimenez2023SWEBench} constructs tasks from real GitHub issues and the corresponding pull requests, requiring models to resolve issues end-to-end by repository edits. Building on this benchmark, SWE-agent demonstrates that designing stronger agent-computer interfaces enables more effective navigation, file editing, and test execution~\BlueNumber{\cite{Yang2024SWEagent}}. OpenHands~\BlueNumber{\cite{Wang2024OpenDevin}} further generalizes this paradigm by providing an open platform for agent evaluation. While these studies significantly advance agent evaluation, they predominantly assume an existing codebase as the starting point, which differs fundamentally from the NL2Repo setting.

More closely related to our work are approaches that explicitly target repository- or project-level generation from scratch. CodeAgent introduces a tool-integrated framework for repo-level code generation and a dedicated benchmark, CodAgentBench~\BlueNumber{\cite{Zhang2024CodeAgent}}. ProjectGen~\BlueNumber{\cite{Zhao2025ProjectGen}} studies project-level generation from requirements through architecture design, skeleton construction, and code completion guided by a structured architecture representation. These systems are more relevant to the goal of our work, but \tool differs in a key aspect: its software design sketch (SDS) serves not merely as a planning artifact, but as a machine-checkable contract that specifies file ownership, public interfaces, and dependency constraints, and that actively governs every subsequent implementation step.

Even in the absence of multi-agent architectures, iterative repository-level generation with intermediate feedback has been shown to improve performance on generation tasks. For example, AlphaCodium~\BlueNumber{\cite{Ridnik2024AlphaCodium}} is a test-driven multi-stage pipeline that outperforms single-pass prompting on code generation tasks. Similarly, Self-Debugging~\cite{Chen2024SelfDebugging} enables LLMs to refine their own generated code by explaining and re-executing it, and Reflexion~\BlueNumber{\cite{Shinn2024Reflexion}} equips agents with verbal self-reflection to improve decision-making across development stages. These observations reinforce an important finding for NL2Repo: when tasks involve many dependent decisions, task decomposition and iterative verification are not just implementation conveniences, but central to generating correct repositories.

\subsection{Conclusive Summary}\label{subsec:conclusive-summary}
Existing literature has addressed several complementary aspects of repository-level generation under varying operating conditions. Early work on NL2Code and long-context modeling primarily improves local synthesis and context utilization. Studies on repository-level completion and feature-implementation benchmarks focus on generation within existing codebases. Meanwhile, agent-based frameworks demonstrate the benefits of role specialization, tool integration, and iterative verification in managing complex generation workflows. These studies suggest that high-quality software generation depends not only on model capacity, but also on how planning, context, and execution are organized across the end-to-end workflow.

While CodeS~\BlueNumber{\cite{Zan2025CodeS}} and NL2Repo-Bench~\BlueNumber{\cite{Ding2025NL2RepoBench}} establish NL2Repo as a meaningful setting for repository-level code generation from natural-language requirements, the broader methodological space for this setting remains relatively underexplored. In particular, unlike tasks defined over an existing repository, NL2Repo introduces additional challenges, including repository planning, cross-file coordination, and consistency maintenance throughout the generation process. To better address these challenges, we propose \tool, a multi-agent framework for end-to-end repository construction in the NL2Repo setting. Built on sketch-based planning and agent collaboration, \tool further strengthens coordination and iterative refinement through SDS competition, \agentrole{cto}-guided contract normalization, dependency-aware scheduling, lightweight Git coordination, and \agentrole{qa}-driven repair mechanisms.

\section{Study Design}\label{sec:study-design}

\subsection{Research Questions}\label{subsec:rqs}
We investigate whether a workflow-oriented multi-agent system improves NL2Repo generation, and  which mechanisms contribute  to  \DIFaddbegin \DIFadd{repository quality}\DIFaddend . To this end, we formulate three research questions (RQs) that progressively evaluate the system at the end-to-end, planning, and implementation levels. RQ1 evaluates controlled end-to-end performance on \sketcheval~\BlueNumber{\cite{Zan2025CodeS}} and uses \nlrepobench ~\BlueNumber{\cite{Ding2025NL2RepoBench}} as an external execution-based validation protocol,  \DIFaddbegin \DIFadd{covering }\DIFaddend all 104 tasks  to  \DIFaddbegin \DIFadd{distinguish structural alignment, partial test passing, and full-task resolution}\DIFaddend . RQ2 focuses on the planning stage,  \DIFaddbegin \DIFadd{examining how }\DIFaddend retrieval-augmented grounding  \DIFaddbegin \DIFadd{affects repository structure}\DIFaddend , and RQ3 analyzes how coordination mechanisms affect repository construction during implementation by scheduling and Git-based communication.

\begin{rqbox}
\textbf{RQ1: How  \DIFaddbegin \DIFadd{does }\DIFaddend CodeTeam's PE  \DIFaddbegin \DIFadd{version }\DIFaddend perform on repository-level code generation tasks compared with CodeS and other baseline methods?}
\end{rqbox}
To answer RQ1, we  compare  \DIFaddbegin \DIFadd{the }\DIFaddend PE  \DIFaddbegin \DIFadd{versions }\DIFaddend of \DIFaddbegin \DIFadd{\tool, }\DIFaddend \codes\BluePunctUnskip{}\DIFaddbegin \DIFadd{, }\DIFaddend and representative agent-based baselines adapted to NL2Repo. We  \DIFaddbegin \DIFadd{evaluate structural quality }\DIFaddend on \sketcheval and complement  \DIFaddbegin \DIFadd{it }\DIFaddend with  execution-based  \DIFaddbegin \DIFadd{evaluation }\DIFaddend on \nlrepobench. \DIFaddbegin \DIFadd{We restrict the comparison to PE because stage-specific SFT supervision cannot be matched consistently across these workflows.
}\DIFaddend 

\begin{rqbox}
\textbf{RQ2: Does retrieval-augmented grounding improve architectural planning in NL2Repo tasks?}
\end{rqbox}
Planning mistakes in NL2Repo (e.g., missing modules or inconsistent interfaces) often propagate to downstream implementation. \tool incorporates an optional RAG subsystem in the architect stage, combined with a competition-and-selection workflow. To isolate its effect, we  \DIFaddbegin \DIFadd{disable }\DIFaddend retrieval while keeping the rest of the workflow intact, and  \DIFaddbegin \DIFadd{examine the resulting structural impact across controlled ablation settings}\DIFaddend .

\begin{rqbox}
\textbf{RQ3: Do dynamic developer allocation and lightweight Git-based coordination affect repository generation quality?}
\end{rqbox}
Repository generation requires managing inter-file dependencies and frequent interface updates. \tool addresses these challenges with \agentrole{Architect}-guided developer allocation, dependency-aware execution, and a Git-based coordination that propagates interface changes under a bounded context. For answering RQ3, we \DIFaddbegin \DIFadd{systematically }\DIFaddend ablate these coordination mechanisms\DIFaddbegin \DIFadd{, evaluating how developer allocation, interface propagation, }\DIFaddend and  \DIFaddbegin \DIFadd{QA-driven repair shape structural coherence }\DIFaddend and  \DIFaddbegin \DIFadd{execution correctness}\DIFaddend .

\DIFaddbegin \DIFadd{Table~\BlueNumber{\ref{tab:rq-evidence}} summarizes the evaluation used to address each RQ; the statistical protocol is described in Section~\BlueNumber{\ref{subsec:metrics}}.
}

\begin{table}[t]
\BlueNewFloat

\centering
\small
\caption{\DIFaddFL{Research questions and the corresponding evaluations.}}\label{tab:rq-evidence}
\setlength{\tabcolsep}{4pt}
\renewcommand{\arraystretch}{1.1}
\begin{tabularx}{\linewidth}{@{}lXl@{}}
\toprule
\textbf{\DIFaddFL{RQ}} & \textbf{\DIFaddFL{Evidence scope}} & \textbf{\DIFaddFL{Results}} \\
\midrule
\DIFaddFL{RQ1 }& \DIFaddFL{Seven PE workflows; structural and execution results; cost and diagnostic analyses }& \DIFaddFL{Section~\ref{subsec:rq1}--\ref{subsec:nl2repo-results} }\\
\DIFaddFL{RQ2 }& \DIFaddFL{Retrieval removal (A5), planning variants, and the architect-count/retrieval-depth grid }& \DIFaddFL{Section~\ref{subsec:rq2} }\\
\DIFaddFL{RQ3 }& \DIFaddFL{Allocation, scheduling, interface propagation, and QA variants; eight execution-subset contrasts }& \DIFaddFL{Section~\ref{subsec:rq3} }\\
\bottomrule
\end{tabularx}
\end{table}
\DIFaddend \subsection{Design and Implementation of \tool}
\label{subsec:implementation}
\begin{figure}[t]
\centering
\includegraphics[width=0.92\linewidth]{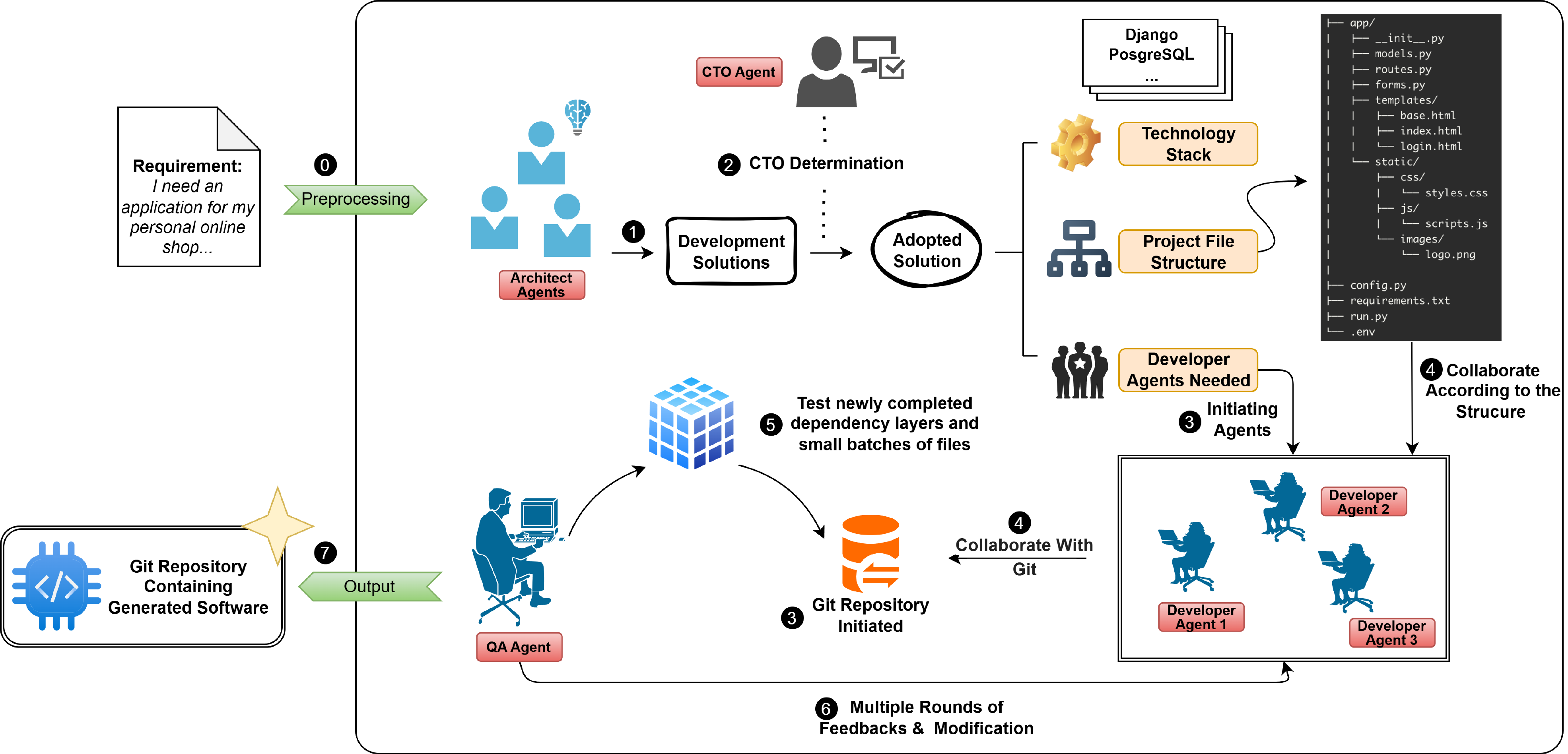}
\caption{\tool workflow from requirements preprocessing (\circledzero) to architect planning (\ding{182}), \agentrole{cto} selection (\ding{183}), solution materialization and Git-coordinated implementation (\ding{184}--\ding{185}), \agentrole{qa}-driven repair (\ding{186}--\ding{187}), and final repository output (\ding{188}).}
\Description{Workflow diagram of CodeTeam}
\label{fig:architecture}
\end{figure}

Figure~\ref{fig:architecture} illustrates the end-to-end workflow of \tool. Starting from a README-style requirements document (the input format used in \sketcheval benchmark~\BlueNumber{\cite{Zan2025CodeS}} and \nlrepobench benchmark~\BlueNumber{\cite{Ding2025NL2RepoBench}}), the system first performs rule-based preprocessing (Step~\circledzero) to produce a normalized requirements document. The workflow then enters a planning-and-selection phase: in Step~\ding{182}, $N$ \agentrole{Architect} agents independently propose alternative software design sketches (SDSs); in Step~\ding{183}, a \agentrole{cto} agent evaluates these candidates, performs comparative analysis, and selects a single adopted solution. The adopted SDS explicitly determines the three design artifacts highlighted in Figure~\ref{fig:architecture}: the technology stack, the project file structure, and the developer allocation plan. In Step~\ding{184}, \tool materializes the adopted SDS by initializing a Git repository, pre-constructing the file hierarchy, and instantiating exactly the \agentrole{Developer} agents specified in the selected developer plan. Step~\ding{185} corresponds to the implementation stage, during which the \agentrole{Developer} agents generate code following the prescribed structure while coordinating through lightweight Git-based communication. In Step~\ding{186}, the \agentrole{qa} agent evaluates newly completed dependency layers or small batches of files against the requirements and the SDS. In Step~\ding{187}, QA findings are routed back to the responsible \agentrole{Developer} agents, and any impacted dependent files are requeued when interface changes propagate. Steps~\ding{185}--\ding{187} repeat until one of the following termination conditions is met: no new issues are detected, a predefined iteration limit is reached, or resource constraints (e.g., wall-clock time or token budget) are exhausted. Finally, Step~\ding{188} outputs the generated Git repository. 

Unless otherwise specified, the default configuration uses $N{=}4$ \agentrole{Architect} agents, one \agentrole{cto} agent, and one \agentrole{qa} agent. The number of \agentrole{Developer} agents is not fixed globally. Instead, each \agentrole{Architect} agent proposes a project-specific developer plan within its SDS, specifying the required team size of \agentrole{Developer} agents, $D$, and a file-to-developer ownership map. After the \agentrole{cto} agent selects an SDS, \tool instantiates exactly those $D$ \agentrole{Developer} agents for the run. 
The choice of agent multiplicity reflects a trade-off between performance and computational cost. Empirically, setting $N{=}4$ \agentrole{Architect} agents provides a favorable balance between design diversity and token efficiency in the current configuration of \tool; \DIFaddbegin \DIFadd{we examine sensitivity to this setting across multiple configurations in Section~\ref{subsec:rq2}}\DIFaddend . %As for the number of \agentrole{Architect} agents, \agentrole{cto} agents, and \agentrole{qa} agents in the current configuration of \tool, it is based on the evaluation and testing of token cost and execution performance. We found that setting the number of \agentrole{Architect} agents to $N{=}4$ provides a good balance between cost and effectiveness.
Meanwhile, \DIFaddbegin \DIFadd{the }\DIFaddend \agentrole{cto} agent and \agentrole{qa} agent are responsible for \DIFaddbegin \DIFadd{centralized selection and verification checkpoints. While candidate contract filtering and queue scheduling follow deterministic rules, comparative SDS scoring and lightweight test synthesis operate as sequential single-pass stages that }\DIFaddend do not substantially benefit from parallelization. Therefore, the current configuration uses a single \agentrole{cto} and a single \agentrole{qa} agent.

To ensure fair evaluation, all agents within a run share the same backbone model and decoding configuration (e.g., temperature, top-$p$, and the remaining sampling parameters). This design isolates the effects of workflow structure and inter-agent coordination, ensuring that observed performance differences arise \DIFaddbegin \DIFadd{primarily }\DIFaddend from the framework itself rather than variations in model behavior\DIFaddbegin \DIFadd{, while cumulative call and token expenditures across workflows are reported explicitly (Section~\BlueNumber{\ref{subsec:config}})}\DIFaddend .

\subsubsection{Requirements Preprocessing}
We treat a repository's README as the input of natural-language requirements. To convert this input into a structured and analysis-ready form, we apply a rule-based preprocessing pipeline. Specifically, we normalize section headings, flatten nested bullet lists, keep fenced code blocks only when they convey actionable information (e.g., commands, configuration keys, or API examples). Nonessential elements (e.g., badges, images, changelogs, release notes, and contributor sections) are removed. The output is a single requirements document with stable section markers. These markers allow subsequent system components to reference requirement spans by heading name rather than by raw line indices. Overall, this preprocessing step keeps requirements-relevant details while eliminating noise that is unlikely to contribute to downstream repository construction.

\subsubsection{Software Design Sketch as an Executable Contract}
\tool represents each candidate plan as a structured \emph{software design sketch} (SDS), a machine-checkable design contract that specifies the repository-level implementation plan before code generation begins. Each SDS records the three design artifacts highlighted in Figure~\ref{fig:architecture}: the technology stack, the project file structure, and the developer allocation plan. Here, the project structure is broader than a directory tree: it includes the fixed repository file hierarchy, each file's responsibility and public interface, and explicit dependency edges among files.

Concretely, an SDS specifies: (1) the technology stack, including programming language, framework, testing framework, and key dependencies; (2) a fixed repository file tree that the system instantiates \emph{before} implementation; (3) per-file responsibilities and public interfaces (e.g., \DIFaddbegin \DIFadd{class }\DIFaddend or \DIFaddbegin \DIFadd{function }\DIFaddend signatures); (4) explicit cross-file dependencies (e.g., imports or calls); (5) a developer allocation plan that determines how many \agentrole{Developer} agents are needed and which files are owned by \textit{Developer~0} to \textit{Developer~$D{-}1$}; and (6) coarse workload estimates to guide execution. By embedding team size and ownership into the SDS, the system treats these factors as project-specific planning decisions rather than fixed global hyperparameters.

From an operational perspective, the SDS must include four mandatory fields: \texttt{tech\_stack}, \texttt{repo\_tree}, \texttt{developer\_plan}, and \texttt{files}. The \texttt{repo\_tree} field records the file hierarchy, while each entry in \texttt{files} records the corresponding file-level contract, including its purpose, exported interface, owner, and dependencies. A candidate is considered \emph{parseable} only if it can be loaded as JSON without manual repair. \emph{Structural validity} then requires that every file listed in \texttt{files} appears in \texttt{repo\_tree}, every owner identifier appears in \texttt{developer\_plan}, every \texttt{depends\_on} edge points to a declared file or exported symbol, and the resulting inter-file dependency graph is acyclic after collapsing same-file symbol references. Candidates that fail these hard checks are still logged for diagnostics, but excluded from \agentrole{cto} selection.

Following plan selection by the \agentrole{cto} agent, \tool initializes a Git repository, creates the full file tree, and instantiates the \agentrole{Developer} agents declared in the chosen SDS. During implementation, developers are constrained to operate strictly within the predefined structure: no extra files outside the SDS can be created, and file ownership must remain consistent with the developer plan. This constraint is \DIFaddbegin \DIFadd{part of the proposed workflow }\DIFaddend to ensure that architectural decisions remain traceable to the planning stage and that comparisons across methods stay reproducible. \DIFaddbegin \DIFadd{Listing}\DIFaddend ~\ref{lst:sds} shows a simplified (illustrative) SDS schema.

\begin{lstlisting}[caption={Simplified SDS schema},label={lst:sds},basicstyle=\ttfamily\scriptsize]
{
  "tech_stack": {"language": "Python", "framework": "Flask", "test": "pytest"},
  "repo_tree": ["src/core.py", "src/api.py", "tests/test_api.py"],
  "dependencies": ["flask>=2.0", "pytest"],
  "developer_plan": {
    "num_developers": 3,
    "owners": [
      {"id": 0, "files": ["src/core.py"]},
      {"id": 1, "files": ["src/api.py"]},
      {"id": 2, "files": ["tests/test_api.py"]}
    ]
  },
  "files": [
    {"path": "src/core.py", "purpose": "domain objects",
     "public_api": ["class User: ..."], "owner": 0},
    {"path": "src/api.py", "purpose": "HTTP API wrappers",
     "public_api": ["def fetch_user(user_id: str) -> User"],
     "depends_on": ["src/core.py::User"], "owner": 1},
    {"path": "tests/test_api.py", "purpose": "API tests",
     "depends_on": ["src/api.py::fetch_user"], "owner": 2}
  ]
}
\end{lstlisting}

\subsubsection{\agentrole{Architect} Stage: Diversified Planning with Optional Retrieval Grounding}
Candidate SDSs are created in the Architect stage. Specifically, \tool instantiates $N$ \agentrole{Architect} agents, each of which independently drafts one candidate SDS. Each candidate specifies not only the repository structure and interfaces, but also a project-specific development plan, including the number of developers to instantiate and the corresponding file ownership assignments. This design enables different architects to explore trade-offs between parallelism and coordination cost. \DIFaddbegin \DIFadd{We use }\DIFaddend $N{=}4$ as \DIFaddbegin \DIFadd{the default }\DIFaddend configuration\DIFaddbegin \DIFadd{; the sensitivity grid for $N$ and $k$ is reported in Section~\ref{subsec:rq2}}\DIFaddend . A common failure mode is design convergence, where different \agentrole{Architect} agents produce highly similar designs, thereby reducing diversity and weakening competitive selection. To mitigate this issue, we introduce several mechanisms. First, the generation order of architects is randomized per \DIFaddbegin \DIFadd{run}\DIFaddend . Second, each \agentrole{Architect} agent is assigned a different design preference (e.g., minimal vs. layered architecture, or class-based vs. functional organization). Third, instead of exposing full prior SDS candidates, each \agentrole{Architect} agent only receives a short summary of previously claimed top-level module names and high-level design intents. Consequently, the non-duplication constraint is enforced at the coarse module-boundary level rather than by exposing full candidate plans to downstream architects.

When RAG is enabled, the system retrieves the top-$k$ design references ($k{=}5$ by default) from a curated corpus of public repositories. Each item is stored as a compact representation (including README summary, file tree, dependency patterns, file-role summaries, and exported-interface hints) rather than large raw code blocks. \agentrole{Architect} agents recognize this information as \emph{design hints}, such as module decompositions, dependency choices, typical file boundaries, and plausible team organization. \agentrole{Architect} agents are instructed to produce an original SDS tailored to the current requirements. Before submission, each \agentrole{Architect} agent performs a lightweight self-validation that mirrors the parsing and validity checks, as well as the proposed developer count and file ownership mapping.

\subsubsection{\agentrole{cto} Stage: Selection and Normalization}
CTO selection consists of two steps: SDS candidate filtering and SDS candidate selection. First, an automatic filtering process removes invalid candidates. Specifically, candidates are discarded when they exhibit JSON parsing failures, missing ownership, unresolved file references, inconsistent symbol declarations, or cyclic dependencies at the file level. Second, the \agentrole{cto} agent evaluates each remaining candidate on four criteria: (1) structural validity, (2) interface consistency, (3) implementability under the stated dependencies and resource constraints, and (4) the appropriateness of the proposed developer count and file ownership relative to the repository scale. Each criterion is scored from 0 to 2, yielding a total score in the range $[0,8]$, and the scores are presented in \agentrole{cto} agent's JSON format output. In the case of ties (i.e., when multiple SDS candidates achieve the same total score), preference is given to candidates with fewer undeclared assumptions and lower cross-module dependency fan-out, as these properties indicate better modularity and robustness.

The selected SDS is then normalized into a machine-actionable representation by canonicalizing paths, deduplicating symbols, resolving naming conflicts, materializing the developer-to-file ownership mapping, and constructing a file-level dependency graph for execution. If normalization fails after selection, the next highest-scoring valid SDS candidate is selected. The normalized SDS becomes the single source of truth for \agentrole{Developer} agents and \agentrole{qa} agents.

\subsubsection{Development Stage: \agentrole{Architect}-planned Allocation with Dependency-aware Execution}
\tool implements repositories at the granularity of \emph{file tasks}, where the task pool is constrained by the developer plan selected in the SDS. Accordingly, Step~\ding{185} in Figure~\ref{fig:architecture} comprises two coupled processes: \agentrole{Developer} agents follow the adopted file structure and developer-to-file ownership mapping, and they synchronize changes through Git-based coordination. Each \agentrole{Developer} agent is assigned ownership of a subset of files, and the number of developers varies across repositories. Once the \agentrole{cto} agent accepts an SDS, that developer-to-file ownership mapping becomes the initial division of labor for the entire execution.

To avoid sharing the entire repository with every \agentrole{Developer} agent, \tool constructs a compact task-specific context consisting of: (1) the SDS fragment relevant to the target file; (2) the current content of the target file; and (3) interface briefs for dependent files, including signatures, invariants, and the latest summarized interface changes. \agentrole{Developer} agents may request at most two additional interface briefs per task, but they do not read the full content of files owned by other agents. Each brief exposes only the depended-upon file path, exported symbols, typed signatures when available, stated invariants, and the latest \texttt{update\_reason} summary. This keeps calls focused and reduces long-context pressure while still enabling cross-file consistency.

Algorithm~\ref{alg:scheduler} describes the execution logic.
It maintains a ready queue based on dependency satisfaction and dispatches each ready file to its recorded owner. Files without prerequisites are prioritized early inside each \agentrole{Developer} agent's ownership set; if an interface change occurs, affected dependent files can be requeued for repair and routed back to their owners. The ready-queue priority is a deterministic ordering rather than a learned or tuned score. Among ready files owned by the same \agentrole{Developer} agent, \tool first chooses the file with the longest downstream dependency chain, since delaying such a file is most likely to block later tasks. If there is a tie, it chooses the file with more direct dependent files; if a tie still remains, it chooses the file that appears earlier in the normalized SDS file list. We denote this ordering compactly as
$\mathsf{prio}(f)=(\mathsf{depth}(f), \mathsf{fanout}(f), -\mathsf{order}(f))$.

\begin{algorithm}[t]
\caption{\agentrole{Architect}-planned developer allocation with dependency-aware execution in \tool.}
\label{alg:scheduler}
\DontPrintSemicolon
\KwIn{Normalized SDS, ownership map $O$, file dependency graph $G{=}(V,E)$}
\KwOut{Repository implementation}
initialize \agentrole{Developer} agents from $O$\;
$C \leftarrow \emptyset$ \tcp*{completed files}
$R \leftarrow \{v \in V \mid \text{deps}(v)=\emptyset\}$\;
compute $\mathsf{prio}(v)$ for each $v \in V$\;
\While{\DIFadd{(}$R \neq \emptyset$ or running tasks exist\DIFadd{) and budget remains}}{
  \ForEach{idle \agentrole{Developer} agent $d_i$}{
    \If{$R \cap O(d_i) \neq \emptyset$}{
      $f \leftarrow \arg\max_{g\in R\cap O(d_i)} \mathsf{prio}(g)$\;
      assign $f$ to developer $d_i$ with compact context\;
      \DIFadd{remove $f$ from $R$ and }mark $f$ as running\;
    }
  }
  \If{a developer finishes $f$}{
    commit changes with a structured message including \texttt{update\_reason}\;
    update the interface index and refresh dependent briefs\;
    remove $f$ from running\;
    $C \leftarrow C \cup \{f\}$\;
    add newly-ready files into $R$ based on $C$\;
  }
  \If{QA trigger holds}{
    run tests; route failures to the owners of suspected source files and re-queue affected files if needed\;
  }
}
\end{algorithm}

\subsubsection{Git-based Coordination, QA Feedback, and Termination}
Steps~\ding{186}--\ding{187} in Figure~\ref{fig:architecture} form a \agentrole{qa}-guided repair loop on top of the Git-tracked workspace. To propagate interface updates without sharing full file contents, \tool uses lightweight Git-based coordination. Each \agentrole{Developer} agent works on an isolated branch and commits changes upon completing a file task. Each commit message includes a structured \texttt{update\_reason} record with four elements: the target file, modified exported symbols, a compatibility note, and affected dependent files. These commit messages serve as the primary communication medium, enabling other agents to update dependent files by using summarized interface information.

The \agentrole{qa} agent evaluates intermediate states during the \agentrole{qa}-guided repair loop without attempting to reconstruct hidden reference tests from the benchmarks (e.g., \DIFaddbegin \DIFadd{\nlrepobench}\DIFaddend ~\BlueNumber{\cite{Ding2025NL2RepoBench}}). Instead, after completing each dependency layer or a small batch of ready files, the \agentrole{qa} agent generates lightweight, temporary tests from the SDS and specified requirements. These tests are generated within a temporary QA directory, executed, and removed before final evaluation, ensuring that \DIFaddbegin \DIFadd{temporary QA files are excluded from }\DIFaddend the final delivered file tree \DIFaddbegin \DIFadd{and its evaluation}\DIFaddend . QA feedback is normalized into issue records containing a symptom category, traceback or assertion message, suspected source files, and the minimal repair scope. The scheduler requeues the smallest affected set of files: by default, only the likely source file is requeued and assigned back to its recorded owner for repair, and dependent files are requeued only when the issue impacts a public API, import path, or shared configuration artifact.

To identify the true error source when failures involve multiple dependent files, the \agentrole{qa} agent applies four diagnostic criteria in sequence: (1) \emph{interface consistency}: the suspected source file must declare all symbols and types referenced by the failing dependent file in its public interface; (2) \emph{dependency directionality}: if the failure traceback originates from a dependent file but the root cause is an undeclared or changed interface in a source file, the source file is flagged as the repair target; (3) \emph{structural validity}: the source file must pass syntactic checks independently before the dependent file is re-evaluated; and (4) \emph{minimal repair scope}: when multiple files exhibit cascading failures, only the earliest file in the dependency chain that violates (1) or (3) is requeued by default, and downstream dependents are requeued only when the issue propagates to a public API, import path, or shared configuration artifact.

Execution terminates under one of three conditions: (1) the \agentrole{qa} agent reports no failures over a complete validation pass, (2) the configured QA-iteration limit is reached, or (3) a global wall-clock or token budget is exceeded. This corresponds to the transition from the iterative loop in Steps~\ding{185}--\ding{187} to the final output in Step~\ding{188}.

\DIFaddbegin \DIFadd{The bounded execution policy directly supports Algorithm~\ref{alg:scheduler}: invalid plans are filtered prior to scheduling, active tasks are removed from $R$ upon dispatch so they cannot be assigned redundantly, and budget exhaustion terminates the attempt by returning the latest repository state. We configure at most three QA rounds per attempt, with early termination triggered if all verification checks pass (analyzed in Section~\ref{subsec:rq3}).}

\DIFadd{By establishing an explicit contract, the SDS limits structural drift, ownership ambiguity, and cascading repair propagation. Fixing the file tree can nevertheless restrict adaptation when an initial design omits an essential module, while strict file-level ownership introduces coordination boundaries. We examine these trade-offs through targeted ablation interventions in Section~\ref{subsec:ablation}.}
\DIFaddend 

\subsection{Benchmark Experiment Design}\label{subsec:benchmark}
We evaluate our approach on \sketcheval, a benchmark introduced by \codes~\BlueNumber{\cite{Zan2025CodeS}}. \sketcheval comprises 19 real-world Python repositories, each paired with requirements documents and reference implementations. Following the official \sketcheval split, we use three difficulty groups: \emph{hard} repositories contain more than 10 Python files or more than 2,500 Python code lines; \emph{medium} repositories contain more than 5 Python files or more than 500 Python code lines but do not meet the hard threshold; and the remaining repositories are classified as \emph{easy}. This yields 5 easy, 8 medium, and 6 hard tasks. For each task, the input is the processed requirements document, and the output is a complete repository generated in an empty workspace. \DIFaddbegin \DIFadd{Table~\BlueNumber{\ref{tab:benchmarks}} summarizes the task coverage of \sketcheval and the complementary execution-based benchmark \nlrepobench. On \sketcheval, we report the composite \emph{structural score}, which compares generated and reference repository sketches in terms of directory hierarchy and file-level syntax skeletons~\BlueNumber{\cite{Zan2025CodeS}}, while complementing this structural score with runtime execution-based testing on \nlrepobench to evaluate functional correctness.}\DIFaddend 

\DIFaddbegin \begin{table}[t]
\BlueNewFloat

\centering
\small
\caption{\DIFaddFL{Benchmarks and evaluation coverage.}}\label{tab:benchmarks}
\setlength{\tabcolsep}{3.5pt}
\renewcommand{\arraystretch}{1.1}
\begin{tabularx}{\linewidth}{@{}XrrrrX@{}}
\toprule
\textbf{\DIFaddFL{Benchmark}} & \textbf{\DIFaddFL{Tasks}} & \textbf{\DIFaddFL{Easy}} & \textbf{\DIFaddFL{Medium}} & \textbf{\DIFaddFL{Hard}} & \textbf{\DIFaddFL{Endpoint}} \\
\midrule
\DIFaddFL{\sketcheval }& \DIFaddFL{19 }& \DIFaddFL{5 }& \DIFaddFL{8 }& \DIFaddFL{6 }& \DIFaddFL{Structural score }\\
\DIFaddFL{\nlrepobench }& \DIFaddFL{104 }& \DIFaddFL{26 }& \DIFaddFL{46 }& \DIFaddFL{32 }& \DIFaddFL{Counted-test pass percentage }\\
\bottomrule
\end{tabularx}
\end{table}
\DIFaddend 

We compare \tool against \DIFaddbegin \DIFadd{six }\DIFaddend baselines that cover single-agent generation, multi-agent software development, generalist autonomous agents, and prior task-specific NL2Repo methods.

\begin{itemize}
  \item \textbf{Vanilla.}
  The Vanilla baseline uses a single agent to generate the entire repository without explicit task decomposition, role collaboration, or cross-file iterative repair. \DIFaddbegin \DIFadd{It }\DIFaddend receives the processed requirements document and basic file creation and read/write capabilities \DIFaddbegin \DIFadd{in a single generation pass}\DIFaddend .

  \item \textbf{ChatDev.}
  ChatDev~\cite{Qian2023ChatDev} is an academic multi-agent software development framework. We adapt its design, coding, and testing role decomposition to the NL2Repo setting while restricting its tools to workspace inspection, file editing, and local test execution. ChatDev serves as the primary academic multi-agent baseline for evaluating whether role-based software development workflows transfer to repository-level code generation from natural language requirements.

  \item \textbf{AutoGPT.}
  AutoGPT~\BlueNumber{\cite{AutoGPT}} is a representative open-source generalist autonomous agent framework. We instantiate it as a single Controller agent that iteratively plans, generates code, inspects the workspace, edits files, and executes local tests. This baseline evaluates whether a generic long-horizon agent loop, without task-specific repository planning mechanisms, is sufficient for NL2Repo.

  \item \textbf{\codes.}
  \codes~\BlueNumber{\cite{Zan2025CodeS}} is a task-specific NL2Repo baseline based on multi-layer sketch generation \DIFaddbegin \DIFadd{(\texttt{RepoSketcher}, \texttt{FileSketcher}, and \texttt{SketchFiller}). We evaluate its prompt-engineering (PE) variant}\DIFaddend . \codes serves as the representative prior NL2Repo baseline and provides a direct comparison against methods designed specifically for repository-level generation.

  \item \textbf{\DIFaddbegin \DIFadd{MetaGPT.}\DIFaddend }
  \DIFaddbegin \DIFadd{MetaGPT~\BlueNumber{\cite{Hong2023MetaGPT}} organizes development through standardized operating procedures (SOPs) and structured handoffs among product, architecture, and engineering roles, broadening the comparison of role-based multi-agent workflows.}\DIFaddend 

  \item \textbf{\DIFaddbegin \DIFadd{DeepSeek-style adapter.}\DIFaddend }
  \DIFaddbegin \DIFadd{The DeepSeek-style adapter implements a coordinator/worker task-dispatch workflow with task decomposition, worker implementation, and integration checking on the same backbone as the other workflows (recorded as \texttt{DeepSeek-Adapter} in the replication artifacts). It evaluates a coordinator/worker architecture rather than a native DeepSeek model.}\DIFaddend 
\end{itemize}

\DIFaddbegin \begin{table}[t]
\BlueNewFloat

\centering
\small
\caption{\DIFaddFL{Generation workflows used in the main comparison.}}\label{tab:protocols}
\setlength{\tabcolsep}{3.5pt}
\renewcommand{\arraystretch}{1.1}
\begin{tabularx}{\linewidth}{@{}p{2.2cm}Xp{2cm}@{}}
\toprule
\textbf{\DIFaddFL{Method}} & \textbf{\DIFaddFL{Workflow structure}} & \textbf{\DIFaddFL{Retrieval}} \\
\midrule
\DIFaddFL{Vanilla }& \DIFaddFL{Single-pass direct generation }& \DIFaddFL{None }\\
\DIFaddFL{ChatDev }& \DIFaddFL{Design--coding--testing dialogue }& \DIFaddFL{None }\\
\DIFaddFL{AutoGPT }& \DIFaddFL{Controller/tool loop }& \DIFaddFL{None }\\
\DIFaddFL{\codes }& \DIFaddFL{RepoSketcher, FileSketcher, and SketchFiller }& \DIFaddFL{None }\\
\DIFaddFL{MetaGPT }& \DIFaddFL{Structured handoffs across SOP roles }& \DIFaddFL{None }\\
\DIFaddFL{DeepSeek-style adapter }& \DIFaddFL{Coordinator/worker task dispatch }& \DIFaddFL{None }\\
\DIFaddFL{\tool }& \DIFaddFL{SDS, CTO, scheduler, and QA }& \DIFaddFL{None (main) }\\
\bottomrule
\end{tabularx}
\end{table}
\DIFaddend 

For a fair comparison, none of the baselines receives SDS generation, architect competition, retrieval hints, dependency-aware developer allocation, Git-style interface specifications, or the ownership-aware QA repair mechanism used by \tool. Within each controlled comparison, all agents are instantiated with the same backbone model and decoding configuration. \DIFaddbegin \DIFadd{In the primary seven-workflow comparison (Table~\BlueNumber{\ref{tab:protocols}}), all systems---including \tool---operate without external retrieval so that the comparison isolates multi-agent planning and coordination. The retrieval-enabled configuration $F$ is evaluated separately in the ablation experiments (Section~\ref{subsec:ablation}). Because cumulative LLM invocations, token throughput, and repair rounds vary across workflow architectures, we report per-workflow resource usage alongside quality metrics.}\DIFaddend 

\subsubsection{Backbone Model and Inference Settings}\DIFaddbegin \label{subsec:config}\DIFaddend 
\DIFaddbegin \DIFadd{All evaluated workflows use Qwen3.8-27B~\BlueNumber{\cite{Qwen2026Qwen38}} with BF16 inference, thinking mode with medium reasoning effort, a temperature of $0.6$, top-$p$ of $0.95$, and at most 8,192 output tokens per call. The primary evaluation is conducted under prompt engineering across five independent repetitions per task, yielding 95 structural runs and 520 execution runs per workflow (with at most three QA rounds per attempt in \tool\BluePunctUnskip{}). Cumulative call and token expenditures across workflows are analyzed in Section~\ref{subsec:nl2repo-results}. We chose Qwen3.8-27B, a recent model (released on August 14, 2026) checkpoint with publicly available weights, because we could host it locally within our compute budget for the expanded evaluation. We used the same checkpoint across workflows to control for backbone differences, enable reproducible accounting of token and compute usage, and avoid hosted API version drift.}\DIFaddend 

\subsubsection{SFT Setup}
\DIFaddbegin \DIFadd{We focus our comparative evaluation on the prompt-engineering (PE) setting to maintain methodological consistency across all seven workflows, because stage-specific }\DIFaddend supervised fine-tuning (SFT) \DIFaddbegin \DIFadd{data cannot be constructed symmetrically across distinct agent architectures. An earlier SFT exploration conducted on }\DIFaddend Qwen2.5-72B~\cite{Qwen2024Qwen25} \DIFaddbegin \DIFadd{(summarized in Table~\BlueNumber{\ref{tab:sft}}) is excluded from the comparative evaluation so that all reported rankings reflect matched PE runs on the shared Qwen3.8-27B backbone.}\DIFaddend 

\DIFaddbegin \begin{table}[t]
\BlueNewFloat

\centering
\small
\caption{\DIFaddFL{Scope of the PE evaluation and the excluded SFT pilot.}}\label{tab:sft}
\setlength{\tabcolsep}{3.5pt}
\renewcommand{\arraystretch}{1.1}
\begin{tabularx}{\linewidth}{@{}p{2.8cm}X@{}}
\toprule
\textbf{\DIFaddFL{Aspect}} & \textbf{\DIFaddFL{Evaluation scope}} \\
\midrule
\DIFaddFL{Current comparison }& \DIFaddFL{PE only; five repetitions per task }\\
\DIFaddFL{Reason for excluding SFT }& \DIFaddFL{Stage-specific supervision is not matched symmetrically across the seven workflows }\\
\DIFaddFL{Prior campaign metadata }& \DIFaddFL{Qwen2.5-72B~\BlueNumber{\cite{Qwen2024Qwen25}}; 18,240 training and 960 validation instances; 4,266 GPU-hours }\\
\DIFaddFL{Current backbone }& \texttt{\DIFaddFL{Qwen3.8-27B}}\DIFaddFL{~\BlueNumber{\cite{Qwen2026Qwen38}}; shared across all PE workflows }\\
\bottomrule
\end{tabularx}
\end{table}
\DIFaddend 

\subsubsection{RAG Corpus and Retrieval Settings}
The \agentrole{Architect}-stage RAG corpus is built from \DIFaddbegin \DIFadd{public }\DIFaddend Python repositories. We exclude benchmark repositories, as well as forks, mirrors, archived projects, boilerplate templates, and any repositories overlapping with training or test targets by name, origin, package lineage, dependency manifests, or high README similarity. For indexing, each repository is converted into chunks rather than raw code, including README summaries, top-level file trees, dependency manifests, and brief file-role and interface summaries. The resulting artifacts are segmented into 768-token chunks with 128-token overlap and embedded using BGE-M3~\cite{Chen2024BGEM3}. We index these embeddings using an HNSW~\cite{Malkov2016HNSW} index implemented in FAISS~\BlueNumber{\cite{Douze2024Faiss}}. At query time, the architect stage retrieves top-$k{=}5$ chunks from distinct repositories based on cosine similarity. This design prioritizes reusable architectural patterns while minimizing the risk of implementation-level copying.

\DIFaddbegin \DIFadd{The corpus contains 1,240 repositories and 186,000 chunks, collected on 2025-11-30. To verify the absence of benchmark leakage, we audited 200 representative retrieval probes against a cosine similarity threshold of $0.72$, observing zero suspicious overlaps (Table~\BlueNumber{\ref{tab:rag-qa}}). We further discuss contamination threats to the external validity in Section~\ref{subsec:threats-external}.}

\BlueSectionNumberBegin \subsubsection{\DIFadd{Statistical Analysis and Reproducibility}}\BlueSectionNumberEnd \label{subsec:metrics}
\DIFadd{To assess differences between workflows, we first average the five repetitions within each task and then compute paired within-task differences (Table~\BlueNumber{\ref{tab:stats}}). We use 20,000 task-bootstrap resamples~\BlueNumber{\cite{Efron1979}} with a fixed seed of 20261030 for the main and structural intervention confidence intervals. We apply two-sided paired $t$ tests with Holm step-down adjustment separately across the five main contrasts, thirteen structural interventions, and eight execution-subset interventions. For the 24-task execution subset, we report 95\% paired $t$ confidence intervals over the 24 task means. Because the task is the unit of both resampling and inference, repeated runs are not treated as independent tasks.}

\begin{table}[t]
\BlueNewFloat

\centering
\small
\caption{\DIFaddFL{Statistical analysis protocol for workflow and intervention comparisons.}}\label{tab:stats}
\setlength{\tabcolsep}{3.5pt}
\renewcommand{\arraystretch}{1.1}
\begin{tabularx}{\linewidth}{@{}p{3cm}X@{}}
\toprule
\textbf{\DIFaddFL{Item}} & \textbf{\DIFaddFL{Protocol}} \\
\midrule
\DIFaddFL{Unit of analysis }& \DIFaddFL{Task means over five repetitions; repetitions are not treated as independent tasks }\\
\DIFaddFL{Paired contrasts }& \DIFaddFL{Within-task differences between the indicated workflows or between an intervention and its reference }\\
\DIFaddFL{Main/structural intervals }& \DIFaddFL{95\% percentile task bootstrap, 20,000 resamples; seed 20261030 }\\
\DIFaddFL{Execution-subset intervals }& \DIFaddFL{95\% paired $t$ intervals over 24 task means }\\
\DIFaddFL{Multiplicity }& \DIFaddFL{Paired $t$ tests with separate Holm families: five reported main contrasts, thirteen structural interventions, and eight execution-subset interventions }\\
\DIFaddFL{Sensitivity }& \DIFaddFL{Leave-one-task-out (LOO) mean ranges across tasks }\\
\DIFaddFL{Auxiliary summaries }& \DIFaddFL{Compute-matched controls and QA diagnostics are summarized descriptively over 520 records }\\
\bottomrule
\end{tabularx}
\end{table}

\DIFadd{Across all experiments, the evaluation comprises 665 main structural runs ($7 \times 19 \times 5$), 1,330 structural intervention runs ($14 \times 19 \times 5$ across reference $F$ and A1--A13), 855 sensitivity-grid runs ($9 \times 19 \times 5$), 3,640 main execution runs ($7 \times 104 \times 5$), and 1,080 execution-subset runs ($9 \times 24 \times 5$), together with compute-control and QA diagnostic records. All run-level result tables, diagnostic summaries, and statistical analysis scripts are provided in our replication package~\BlueNumber{\cite{replpack}}.}
\DIFaddend 

\subsection{Execution-Based Evaluation on \nlrepobench}\label{subsec:nl2repobench}
To assess whether the conclusions on \sketcheval generalize to a stricter execution-based setting, we introduce an external evaluation-based protocol on \nlrepobench~\BlueNumber{\cite{Ding2025NL2RepoBench}}. \nlrepobench contains 104 tasks spanning nine categories of Python libraries. Each task provides the agent with a single natural-language specification and an empty workspace, and the generated repository is evaluated by executing the original upstream \texttt{pytest} suite in a controlled environment~\BlueNumber{\cite{Ding2025NL2RepoBench}}. The benchmark groups tasks by repository size into \textit{easy} (26), \textit{medium} (46), and \textit{hard} (32), with an average input length of approximately 18.8k tokens~\BlueNumber{\cite{Ding2025NL2RepoBench}}. Compared to \sketcheval, \nlrepobench emphasizes end-to-end executability: generated repositories must satisfy package layout, dependency specification, import consistency, and functional correctness as verified by the upstream test suites~\BlueNumber{\cite{Ding2025NL2RepoBench}}.

In our work, \nlrepobench is used as an external validation protocol. We follow the document-only setting of \nlrepobench, where agents do not have access to the target repository, scaffolding, or test cases during generation. A practical adaptation is needed because \tool decomposes repository construction into architect competition, CTO-based sketch selection, developer implementation, Git-mediated coordination, and QA-driven repair. Crucially, this adaptation affects only how the repository is generated \DIFaddbegin \DIFadd{while preserving the benchmark input condition and final upstream evaluation. Every workflow in Table~\BlueNumber{\ref{tab:nl2repo-main}} is evaluated on all 104 tasks across five repetitions.}\DIFaddend 

\DIFaddbegin \DIFadd{We report three execution outcomes on \nlrepobench: (1) the \emph{mean counted-test pass percentage} ($100\,n_{\mathrm{passed}}/n_{\mathrm{total}}$), averaged across the five repetitions within each task and then across tasks; (2) \emph{harness-clean} attempts, defined as runs in which all collected tests pass with zero recorded test failures ($n_{\mathrm{failed}}{=}0$, $n_{\mathrm{passed}}{=}n_{\mathrm{collected}}$), even if only a subset of the upstream suite is collected; and (3) \emph{full-task resolution} (\PassAtOne{}), which requires 100\% of upstream counted tests to pass ($n_{\mathrm{passed}}{=}n_{\mathrm{total}}$) without collection or execution errors. Across all 520 attempts per workflow, no method achieves full-task resolution (\PassAtOne{}${=}0\%$), reflecting the difficulty of from-scratch repository generation under unmodified upstream test suites.}

\DIFadd{Test collection statistics are analyzed alongside the execution results in Section~\ref{subsec:nl2repo-results}. For execution-based ablations, we evaluate eight representative interventions and reference $F$ on a stratified 24-task subset across five repetitions per configuration in Section~\ref{subsec:rq3}.}
\DIFaddend 

\subsection{Ablation Study Design}\label{subsec:ablation}\DIFaddbegin \label{subsec:interventions}\DIFaddend 
To answer RQ2 and RQ3, we design a set of ablation variants \DIFaddbegin \DIFadd{(A1--A13, summarized in Table~\BlueNumber{\ref{tab:ablation-variants}}) evaluated relative to a common retrieval-enabled reference configuration $F$ ($N{=}4$ architects, top-$k{=}5$ retrieval, mean structural score $55.90$). Reference $F$ serves as the paired baseline for the thirteen interventions and is kept distinct from the retrieval-free main comparison.}\DIFaddend 

\begin{table}[t]
\BlueCaptionNumber

\centering
\small
\caption{\DIFaddbeginFL \DIFaddFL{Definitions and scope of the thirteen interventions.}\DIFaddendFL }\label{tab:ablation-variants}
\DIFaddbeginFL \setlength{\tabcolsep}{3.5pt}
\renewcommand{\arraystretch}{1.1}
\begin{tabularx}{\linewidth}{@{}lXp{2cm}@{}}
\toprule
\textbf{\DIFaddFL{ID}} & \textbf{\DIFaddFL{Change from reference $F$}} & \textbf{\DIFaddFL{Scope}} \\
\midrule
\DIFaddFL{A1 }& \DIFaddFL{One Architect, with the saved planning allowance reallocated to implementation. }& \DIFaddFL{Bundled }\\
\DIFaddFL{A2 }& \DIFaddFL{Four sketches from the same prompt, retaining CTO selection. }& \DIFaddFL{Single policy }\\
\DIFaddFL{A3 }& \DIFaddFL{Select the first structurally valid sketch without comparative CTO ranking. }& \DIFaddFL{Single policy }\\
\DIFaddFL{A4 }& \DIFaddFL{Replace the SDS-governed workflow with a direct implementation plan. }& \DIFaddFL{Bundled }\\
\DIFaddFL{A5 }& \DIFaddFL{Disable Architect-stage retrieval ($k=0$). }& \DIFaddFL{Single policy }\\
\DIFaddFL{A6 }& \DIFaddFL{Fix the developer count at $D=4$ and re-derive the ownership map. }& \DIFaddFL{Bundled }\\
\DIFaddFL{A7 }& \DIFaddFL{Assign files round-robin, changing file grouping and repair ownership. }& \DIFaddFL{Bundled }\\
\DIFaddFL{A8 }& \DIFaddFL{Disable dependent-file regeneration following interface changes. }& \DIFaddFL{Single policy }\\
\DIFaddFL{A9 }& \DIFaddFL{Provide full dependent source files instead of compact interface briefs. }& \DIFaddFL{Single policy }\\
\DIFaddFL{A10 }& \DIFaddFL{Dispatch ready files lexically instead of by dependency depth and fan-out. }& \DIFaddFL{Single policy }\\
\DIFaddFL{A11 }& \DIFaddFL{Use a shared working tree without agent branches; retain live briefs. }& \DIFaddFL{Single policy }\\
\DIFaddFL{A12 }& \DIFaddFL{Freeze interface briefs; omit updates from }\texttt{\DIFaddFL{update\_reason}} \DIFaddFL{records. }& \DIFaddFL{Single policy }\\
\DIFaddFL{A13 }& \DIFaddFL{Disable QA-driven testing and source repair. }& \DIFaddFL{Single policy }\\
\bottomrule
\end{tabularx}
\DIFaddendFL \end{table}

\DIFaddbegin \DIFadd{These interventions distinguish isolated single-policy modifications (A2, A3, A5, and A8--A13) from bundled configuration changes (A1, A4, A6, and A7) where altering an upstream structural decision simultaneously affects downstream budget allocation, contract governance, or file ownership.}
\DIFaddend 

\subsubsection{Ablation for RQ2: Disabling RAG in the \agentrole{Architect} Stage}\label{subsubsec:ablation-rag}
To evaluate the role of retrieval, \DIFaddbegin \DIFadd{variant A5 }\DIFaddend removes the retrieval subsystem entirely. \agentrole{Architect} agents receive only the requirements document and the competition constraints, without access to external references. The \agentrole{cto} agent still selects and normalizes an SDS, and the downstream \agentrole{Developer} agents and \agentrole{qa}-guided repair loop remain unchanged. \DIFaddbegin \DIFadd{We evaluate A5 on all 19 \sketcheval tasks alongside the $3\times 3$ sensitivity grid over architect count $N$ and retrieval depth $k$.}\DIFaddend 

\subsubsection{Ablation for RQ3: Disabling Dynamic Allocation in the \agentrole{Developer} Stage}\label{subsubsec:ablation-dynalloc}
To assess the impact of dynamic developer allocation, we \DIFaddbegin \DIFadd{separate fixed team sizing (A6, fixing $D{=}4$ with re-derived ownership) from round-robin file assignment (A7, distributing files round-robin across developers and altering repair ownership). We additionally evaluate disabling dependent-file regeneration (A8), providing full dependent source files instead of compact briefs (A9), and dispatching ready files lexically rather than by dependency priority (A10).}\DIFaddend 

\subsubsection{Ablation on Git-based Coordination}\label{subsubsec:ablation-git}
To assess the impact of commit-based coordination, we \DIFaddbegin \DIFadd{separate removing isolated agent branches while retaining live briefs in a shared working tree (A11) from freezing interface briefs to their initial post-planning state without propagating \texttt{update\_reason} updates (A12). Finally, A13 disables QA-driven testing and source repair.}\DIFaddend 

\DIFaddbegin \DIFadd{All thirteen interventions are evaluated on the 19 \sketcheval tasks across five repetitions per configuration. Eight representative interventions (A1, A3, A4, A6, A10, A11, A12, and A13) together with reference $F$ are also evaluated on the 24-task \nlrepobench subset (Table~\BlueNumber{\ref{tab:exec-subset}}).}\DIFaddend 

\section{Results}\label{sec:results}
This section reports the experimental results corresponding to the three RQs formulated in Section~\ref{subsec:rqs}, following the evaluation protocol described in Section~\ref{subsec:benchmark} to Section~\ref{subsec:ablation}. We first present the \DIFaddbegin \DIFadd{structural evaluation }\DIFaddend on \sketcheval (RQ1), followed by execution-based validation on \nlrepobench. We then analyze the \DIFaddbegin \DIFadd{intervention contrasts }\DIFaddend that address planning-stage retrieval (RQ2) and implementation-stage coordination (RQ3). Unless stated otherwise, \DIFaddbegin \DIFadd{task scores }\DIFaddend are averaged over \DIFaddbegin \DIFadd{five repetitions}\DIFaddend . In addition to aggregate scores, we examine task-level differences across the 19 \sketcheval repositories to assess whether improvements are broadly distributed rather than dominated by a few outlier tasks.

\subsection{RQ1: \sketcheval Performance of CodeTeam}\label{subsec:rq1}\DIFaddbegin \label{subsec:main-results}\DIFaddend 
Table~\BlueNumber{\ref{tab:rq1-main}} reports the main benchmark results on \sketcheval.

\subsubsection{Overall Performance in \DIFaddbegin \DIFadd{the }\DIFaddend PE Setting}\label{subsubsec:rq1-overview}
In the prompt-engineering (PE) setting, \tool achieves an overall \DIFaddbegin \DIFadd{structural score of 53.88, compared with 50.04 for \codes, an improvement of 3.84 points. The DeepSeek-style adapter is the closest comparator at 52.61, while MetaGPT, ChatDev, AutoGPT, and Vanilla obtain mean scores of 48.51, 47.62, 46.83, and 42.88, respectively. The ablations in Section~\ref{subsec:rq2} and Section~\ref{subsec:rq3} examine which workflow policies contribute to these differences.}\DIFaddend 

A closer look at the per-difficulty breakdown \DIFaddbegin \DIFadd{shows how performance varies across repository complexity levels. On \emph{easy} repositories, \tool and \codes achieve mean scores of 57.09 and 51.44 (+5.65 points); on \emph{medium} repositories, 54.06 and 50.00 (+4.06 points); and on \emph{hard} repositories, 50.98 and 48.93 (+2.05 points). The DeepSeek-style adapter reaches 51.44 on hard repositories compared with 50.98 for \tool\BluePunctUnskip{}. While \tool maintains a positive margin over \codes across all three difficulty groups, the margin narrows on harder repositories as inter-module dependencies and coordination demands increase.}\DIFaddend 

\DIFaddbegin \begin{table}[t]
\BlueCaptionNumber

\centering
\small
\caption{\DIFaddFL{Structural scores in the PE setting for all 19 \sketcheval tasks (five repetitions per task; 95 records per workflow). Overall values report mean$\pm$sample SD across the 95 records.}}\label{tab:rq1-main}\label{tab:main}\label{tab:difficulty}
\setlength{\tabcolsep}{3.5pt}
\renewcommand{\arraystretch}{1.1}
\begin{tabularx}{\linewidth}{@{}Xrrrr@{}}
\DIFaddendFL \toprule
\textbf{Method} & \textbf{Easy (5)} & \textbf{Medium (8)} & \textbf{Hard (6)} & \textbf{All (19)} \\
\midrule
Vanilla & \DIFaddbeginFL \DIFaddFL{47.13 }\DIFaddendFL & \DIFaddbeginFL \DIFaddFL{42.75 }\DIFaddendFL & \DIFaddbeginFL \DIFaddFL{39.54 }\DIFaddendFL & \DIFaddbeginFL \DIFaddFL{42.88{\scriptsize$\pm$7.31} }\DIFaddendFL \\
ChatDev & \DIFaddbeginFL \DIFaddFL{50.01 }\DIFaddendFL & \DIFaddbeginFL \DIFaddFL{47.67 }\DIFaddendFL & \DIFaddbeginFL \DIFaddFL{45.58 }\DIFaddendFL & \DIFaddbeginFL \DIFaddFL{47.62{\scriptsize$\pm$7.12} }\DIFaddendFL \\
AutoGPT & \DIFaddbeginFL \DIFaddFL{48.94 }\DIFaddendFL & \DIFaddbeginFL \DIFaddFL{45.60 }\DIFaddendFL & \DIFaddbeginFL \DIFaddFL{46.72 }\DIFaddendFL & \DIFaddbeginFL \DIFaddFL{46.83{\scriptsize$\pm$7.35} }\DIFaddendFL \\
\codes & \DIFaddbeginFL \DIFaddFL{51.44 }\DIFaddendFL & \DIFaddbeginFL \DIFaddFL{50.00 }\DIFaddendFL & \DIFaddbeginFL \DIFaddFL{48.93 }\DIFaddendFL & \DIFaddbeginFL \DIFaddFL{50.04{\scriptsize$\pm$5.70} }\DIFaddendFL \\
\DIFaddbeginFL \DIFaddFL{MetaGPT }& \DIFaddFL{52.03 }& \DIFaddFL{48.74 }& \DIFaddFL{45.28 }& \DIFaddFL{48.51{\scriptsize$\pm$7.12} }\\
\DIFaddFL{DeepSeek-style adapter }& \DIFaddFL{55.65 }& \DIFaddFL{51.60 }& \DIFaddFL{51.44 }& \DIFaddFL{52.61{\scriptsize$\pm$5.80} }\\
\DIFaddFL{\tool }& \DIFaddFL{57.09 }& \DIFaddFL{54.06 }& \DIFaddFL{50.98 }& \DIFaddFL{53.88{\scriptsize$\pm$6.16} }\\
\DIFaddendFL \bottomrule
\DIFaddbeginFL \end{tabularx}
\end{table}
\DIFaddend 

\DIFaddbegin \DIFadd{Table~\BlueNumber{\ref{tab:contrasts}} reports paired within-task contrasts across the primary comparators. The structural improvement of \tool over \codes (+3.84 points) is statistically significant, with a 95\% task-bootstrap confidence interval of $[2.07, 5.59]$ and $p_{\mathrm{Holm}} = 0.003$. Relative to the DeepSeek-style adapter, \tool shows a $+1.27$-point difference whose confidence interval ($[-0.19, 2.67]$) spans zero ($p_{\mathrm{Holm}} = 0.314$), indicating that the difference between these two coordinated workflows is not statistically significant on the 19-task structural benchmark.}\DIFaddend 

\DIFaddbegin \begin{table}[t]
\BlueNewFloat

\centering
\small
\caption{\DIFaddFL{Paired workflow contrasts (\tool minus comparator), with 95\% task-bootstrap confidence intervals and Holm-adjusted $p$-values across the five primary contrasts.}}\label{tab:contrasts}
\setlength{\tabcolsep}{3.5pt}
\renewcommand{\arraystretch}{1.1}
\begin{tabularx}{\linewidth}{@{}lXrrr@{}}
\toprule
\textbf{\DIFaddFL{Comparator}} & \textbf{\DIFaddFL{Metric}} & \DIFaddFL{$\mathbf{\Delta}$ }& \textbf{\DIFaddFL{95\% CI}} & \DIFaddFL{$\mathbf{p_{Holm}}$ }\\
\midrule
\DIFaddFL{\codes }& \DIFaddFL{Structural }& \DIFaddFL{+3.84 }& \DIFaddFL{$[2.07,5.59]$ }& \DIFaddFL{$0.003$ }\\
\DIFaddFL{\codes }& \DIFaddFL{Counted-test (pp) }& \DIFaddFL{+4.99 }& \DIFaddFL{$[2.19,7.77]$ }& \DIFaddFL{$0.003$ }\\
\DIFaddFL{\codes }& \DIFaddFL{Harness-clean (pp) }& \DIFaddFL{+3.46 }& \DIFaddFL{$[-1.54,8.46]$ }& \DIFaddFL{$0.372$ }\\
\DIFaddFL{Adapter }& \DIFaddFL{Structural }& \DIFaddFL{+1.27 }& \DIFaddFL{$[-0.19,2.67]$ }& \DIFaddFL{$0.314$ }\\
\DIFaddFL{Adapter }& \DIFaddFL{Counted-test (pp) }& \DIFaddFL{+1.67 }& \DIFaddFL{$[-1.03,4.35]$ }& \DIFaddFL{$0.372$ }\\
\bottomrule
\end{tabularx}
\end{table}
\DIFaddend 

\subsubsection{\DIFaddbegin \DIFadd{Scope of the Structural Score}\DIFaddend }\label{subsubsec:rq1-decomp}
\DIFaddbegin \DIFadd{On \sketcheval, the structural score measures repository-level alignment in directory hierarchy, module boundaries, and file-level syntax skeletons~\BlueNumber{\cite{Zan2025CodeS}}. Higher structural scores indicate that a workflow more consistently preserves the target module organization, but structural similarity alone does not verify runtime executability; we therefore pair this evaluation with execution-based testing on \nlrepobench in Section~\ref{subsec:nl2repo-results}.}\DIFaddend 

\subsubsection{Repository Size and Task-level Distribution}\label{subsec:task-level}
Finally, we examine whether the observed improvements are broadly distributed across tasks rather than driven by a few outliers. \DIFaddbegin \DIFadd{As summarized in Table~\BlueNumber{\ref{tab:task-level}}, \tool achieves a higher structural score than \codes}\DIFaddend~on 15 out of 19 repositories, ties on \DIFaddbegin \DIFadd{1 }\DIFaddend (difference within 0.5 points), and scores lower on \DIFaddbegin \DIFadd{3. Task-level differences range from $-4.81$ to $+12.72$ points (median $+4.39$ points), and a leave-one-task-out (LOO) analysis yields mean improvements in $[3.35, 4.32]$ points.}\DIFaddend 

\begin{table}[t]
\BlueCaptionNumber

\centering
\small
\caption{Task-level \DIFaddbeginFL \DIFaddFL{comparisons (\tool minus \codes\BluePunctUnskip{}). Structural ties have }\DIFaddendFL absolute difference \DIFaddbeginFL \DIFaddFL{at most 0.5 points; other ties are exact. LOO denotes the leave-one-task-out mean range.}\DIFaddendFL }\DIFaddbeginFL \label{tab:task-level}\DIFaddendFL \label{tab:rq1-winloss}
\DIFaddbeginFL \setlength{\tabcolsep}{3.5pt}
\renewcommand{\arraystretch}{1.1}
\begin{tabularx}{\linewidth}{@{}Xrrrr@{}}
\toprule
\textbf{\DIFaddFL{Endpoint}} & \textbf{\DIFaddFL{W/T/L}} & \textbf{\DIFaddFL{Range}} & \textbf{\DIFaddFL{Median}} & \textbf{\DIFaddFL{LOO range}} \\
\midrule
\DIFaddFL{Structural }& \DIFaddFL{15/1/3 }& \DIFaddFL{$[-4.81,12.72]$ }& \DIFaddFL{4.39 }& \DIFaddFL{$[3.35,4.32]$ }\\
\DIFaddFL{Counted-test (pp) }& \DIFaddFL{67/1/36 }& \DIFaddFL{$[-36.21,36.84]$ }& \DIFaddFL{5.02 }& \DIFaddFL{$[4.68,5.39]$ }\\
\DIFaddFL{Harness-clean (pp) }& \DIFaddFL{28/54/22 }& \DIFaddFL{$[-80.00,80.00]$ }& \DIFaddFL{0.00 }& \DIFaddFL{$[2.72,4.27]$ }\\
\bottomrule
\end{tabularx}
\DIFaddendFL \end{table}

\DIFaddbegin \begin{table}[t]
\BlueNewFloat

\centering
\small
\caption{\DIFaddFL{Per-task structural scores on \sketcheval, averaged over five repetitions. Differences are computed before rounding.}}\label{tab:per-task}
\setlength{\tabcolsep}{3.5pt}
\renewcommand{\arraystretch}{1.1}
\begin{tabularx}{\linewidth}{@{}rXrrr@{}}
\toprule
\textbf{\DIFaddFL{Task}} & \textbf{\DIFaddFL{Stratum}} & \textbf{\DIFaddFL{CodeS}} & \textbf{\DIFaddFL{CodeTeam}} & \DIFaddFL{$\mathbf{\Delta}$ }\\
\midrule
\DIFaddFL{1 }& \DIFaddFL{Easy }& \DIFaddFL{57.21 }& \DIFaddFL{62.56 }& \DIFaddFL{+5.36 }\\
\DIFaddFL{2 }& \DIFaddFL{Easy }& \DIFaddFL{47.12 }& \DIFaddFL{59.84 }& \DIFaddFL{+12.72 }\\
\DIFaddFL{3 }& \DIFaddFL{Easy }& \DIFaddFL{56.90 }& \DIFaddFL{57.39 }& \DIFaddFL{+0.49 }\\
\DIFaddFL{4 }& \DIFaddFL{Easy }& \DIFaddFL{45.41 }& \DIFaddFL{46.11 }& \DIFaddFL{+0.70 }\\
\DIFaddFL{5 }& \DIFaddFL{Easy }& \DIFaddFL{50.56 }& \DIFaddFL{59.54 }& \DIFaddFL{+8.98 }\\
\DIFaddFL{6 }& \DIFaddFL{Medium }& \DIFaddFL{54.42 }& \DIFaddFL{55.35 }& \DIFaddFL{+0.93 }\\
\DIFaddFL{7 }& \DIFaddFL{Medium }& \DIFaddFL{46.82 }& \DIFaddFL{53.01 }& \DIFaddFL{+6.19 }\\
\DIFaddFL{8 }& \DIFaddFL{Medium }& \DIFaddFL{46.66 }& \DIFaddFL{52.54 }& \DIFaddFL{+5.88 }\\
\DIFaddFL{9 }& \DIFaddFL{Medium }& \DIFaddFL{48.17 }& \DIFaddFL{52.56 }& \DIFaddFL{+4.39 }\\
\DIFaddFL{10 }& \DIFaddFL{Medium }& \DIFaddFL{53.20 }& \DIFaddFL{57.50 }& \DIFaddFL{+4.31 }\\
\DIFaddFL{11 }& \DIFaddFL{Medium }& \DIFaddFL{48.79 }& \DIFaddFL{47.35 }& \DIFaddFL{-1.43 }\\
\DIFaddFL{12 }& \DIFaddFL{Medium }& \DIFaddFL{40.10 }& \DIFaddFL{45.59 }& \DIFaddFL{+5.50 }\\
\DIFaddFL{13 }& \DIFaddFL{Medium }& \DIFaddFL{61.88 }& \DIFaddFL{68.56 }& \DIFaddFL{+6.68 }\\
\DIFaddFL{14 }& \DIFaddFL{Hard }& \DIFaddFL{42.84 }& \DIFaddFL{49.03 }& \DIFaddFL{+6.19 }\\
\DIFaddFL{15 }& \DIFaddFL{Hard }& \DIFaddFL{42.82 }& \DIFaddFL{44.93 }& \DIFaddFL{+2.10 }\\
\DIFaddFL{16 }& \DIFaddFL{Hard }& \DIFaddFL{52.89 }& \DIFaddFL{52.08 }& \DIFaddFL{-0.81 }\\
\DIFaddFL{17 }& \DIFaddFL{Hard }& \DIFaddFL{55.42 }& \DIFaddFL{50.61 }& \DIFaddFL{-4.81 }\\
\DIFaddFL{18 }& \DIFaddFL{Hard }& \DIFaddFL{49.23 }& \DIFaddFL{55.75 }& \DIFaddFL{+6.52 }\\
\DIFaddFL{19 }& \DIFaddFL{Hard }& \DIFaddFL{50.37 }& \DIFaddFL{53.46 }& \DIFaddFL{+3.09 }\\
\bottomrule
\end{tabularx}
\end{table}
\DIFaddend 

\DIFaddbegin \DIFadd{Table~\BlueNumber{\ref{tab:per-task}} reports per-task scores across the 19 repositories. On Task~2 (\textit{easy}), \tool achieves 59.84 compared with 47.12 for \codes (+12.72 points), whereas on Task~17 (\textit{hard}), \tool scores 50.61 versus 55.42 for \codes ($-4.81$ points). These per-task results show that while \tool improves structural alignment on most tasks, repositories with dense inter-file dependencies can still favor simpler sketch generation.}\DIFaddend 

\subsection{RQ1: Execution-Based Validation on \nlrepobench}\label{subsec:nl2repo-results}
To examine whether the improvements observed on \sketcheval generalize to a stricter evaluation setting, we additionally compare methods on \nlrepobench~\BlueNumber{\cite{Ding2025NL2RepoBench}}. In this benchmark, each generated repository is judged by executing the original upstream \texttt{pytest} suite rather than by structural overlap.

Table~\BlueNumber{\ref{tab:nl2repo-main}} reports the results. \DIFaddbegin \DIFadd{\tool achieves a mean counted-test pass rate of 48.24\%, compared with 43.25\% for \codes and 46.57\% for the DeepSeek-style adapter. The paired within-task improvement over \codes is $+4.99$ percentage points (95\% CI $[2.19, 7.77]$, $p_{\mathrm{Holm}} = 0.003$; Table~\BlueNumber{\ref{tab:contrasts}}). Relative to the DeepSeek-style adapter, the $+1.67$ percentage-point difference is not statistically significant (95\% CI $[-1.03, 4.35]$, $p_{\mathrm{Holm}} = 0.372$).}\DIFaddend 

\DIFaddbegin \begin{table}[t]
\BlueCaptionNumber

\DIFaddendFL \centering
\DIFaddbeginFL \footnotesize
\caption{\DIFaddFL{Counted-test pass percentage on \nlrepobench (104 tasks; five repetitions per task). Overall SD is computed across the 520 attempts. A harness-clean attempt has zero recorded failures among collected tests ($n_{\mathrm{failed}}{=}0$, $n_{\mathrm{passed}}{=}n_{\mathrm{collected}}$).}}\label{tab:nl2repo-main}
\setlength{\tabcolsep}{3.5pt}
\renewcommand{\arraystretch}{1.1}
\begin{tabularx}{\linewidth}{@{}Xrrrrr@{}}
\DIFaddendFL \toprule
\textbf{Method} & \textbf{Overall} & \textbf{Easy (26)} & \textbf{Medium (46)} & \textbf{Hard (32)} \DIFaddbeginFL & \textbf{\DIFaddFL{Harness-clean}} \DIFaddendFL \\
\midrule
Vanilla & \DIFaddbeginFL \DIFaddFL{33.34{\scriptsize$\pm$13.92} }& \DIFaddFL{36.90 }& \DIFaddFL{33.21 }& \DIFaddFL{30.64 }& \DIFaddFL{23/520 }\DIFaddendFL \\
ChatDev & \DIFaddbeginFL \DIFaddFL{39.49{\scriptsize$\pm$13.54} }& \DIFaddFL{39.56 }& \DIFaddFL{40.90 }& \DIFaddFL{37.40 }& \DIFaddFL{38/520 }\DIFaddendFL \\
AutoGPT & \DIFaddbeginFL \DIFaddFL{38.92{\scriptsize$\pm$13.86} }& \DIFaddFL{39.77 }& \DIFaddFL{39.83 }& \DIFaddFL{36.93 }& \DIFaddFL{36/520 }\DIFaddendFL \\
\codes & \DIFaddbeginFL \DIFaddFL{43.25{\scriptsize$\pm$13.77} }& \DIFaddFL{44.49 }& \DIFaddFL{43.71 }& \DIFaddFL{41.59 }& \DIFaddFL{67/520 }\\
\DIFaddFL{MetaGPT }& \DIFaddFL{40.74{\scriptsize$\pm$13.19} }& \DIFaddFL{42.53 }& \DIFaddFL{41.12 }& \DIFaddFL{38.72 }& \DIFaddFL{44/520 }\\
\DIFaddFL{DeepSeek-style adapter }& \DIFaddFL{46.57{\scriptsize$\pm$13.33} }& \DIFaddFL{47.51 }& \DIFaddFL{46.32 }& \DIFaddFL{46.17 }& \DIFaddFL{76/520 }\\
\DIFaddFL{\tool }& \DIFaddFL{48.24{\scriptsize$\pm$18.18} }& \DIFaddFL{49.68 }& \DIFaddFL{48.84 }& \DIFaddFL{46.21 }& \DIFaddFL{85/520 }\DIFaddendFL \\
\bottomrule
\DIFaddbeginFL \end{tabularx}
\end{table}
\DIFaddend 

\DIFaddbegin \DIFadd{Across the 104 execution tasks, \tool achieves higher counted-test pass rates than \codes on 67 tasks, ties on 1, and records lower pass rates on 36 (Table~\BlueNumber{\ref{tab:task-level}}). Task-level differences span from $-36.21$ to $+36.84$ percentage points (median $+5.02$ pp; LOO range $[4.68, 5.39]$ pp). For harness-clean attempts, \tool records 85/520 compared with 67/520 for \codes (+3.46 pp; 95\% CI $[-1.54, 8.46]$ pp, $p_{\mathrm{Holm}} = 0.372$).}

\BlueSectionNumberBegin \subsubsection{\DIFadd{Test Coverage and Full-Task Resolution}}\BlueSectionNumberEnd 
\DIFadd{To contextualize these execution scores, we examine upstream test counts and collection coverage. The total number of upstream counted tests ($n_{\mathrm{total}}$) is fixed within each task and ranges from 10 to 80 across the 104 tasks (median 24.5, IQR 19--32). In \tool, collected test cases ($n_{\mathrm{collected}}$) range from 4 to 58 per attempt (median 16), with zero harness-level collection errors (\texttt{collection\_error}${=}0$) recorded across runs.}

\DIFadd{Whereas the counted-test pass rate ($100\,n_{\mathrm{passed}}/n_{\mathrm{total}}$) measures partial test-suite completion relative to the full upstream test count, full-task resolution (\PassAtOne{}) requires $n_{\mathrm{passed}} = n_{\mathrm{total}}$ on a single generation attempt. Across all seven workflows, no attempt passes 100\% of counted tests, yielding an observed \PassAtOne{} of 0/520 for every workflow (Table~\BlueNumber{\ref{tab:pass}}). \tool produces 16 near-complete attempts with counted-test pass rates of at least 90\% and a maximum pass rate of 96.88\%. Among \tool's 85 harness-clean attempts ($n_{\mathrm{failed}}{=}0$, $n_{\mathrm{passed}}{=}n_{\mathrm{collected}}$), the counted-test pass rate averages 64.16\% (ranging from 36.36\% to 95.65\%), indicating that passing all collected tests reflects partial test-suite completion on the collected subset rather than complete task resolution.}

\begin{table}[t]
\BlueNewFloat

\centering
\small
\caption{\DIFaddFL{Full-task resolution and near-complete attempts (520 attempts per workflow). No attempt passes all counted tests; observed \PassAtOne{} is therefore zero for every workflow.}}\label{tab:pass}
\setlength{\tabcolsep}{3.5pt}
\renewcommand{\arraystretch}{1.1}
\begin{tabularx}{\linewidth}{@{}Xrrr@{}}
\toprule
\textbf{\DIFaddFL{Method}} & \textbf{\DIFaddFL{Fully solved}} & \textbf{\DIFaddFL{At least 90\%}} & \textbf{\DIFaddFL{Maximum (\%)}} \\
\midrule
\DIFaddFL{Vanilla }& \DIFaddFL{0 }& \DIFaddFL{0 }& \DIFaddFL{76.47 }\\
\DIFaddFL{ChatDev }& \DIFaddFL{0 }& \DIFaddFL{0 }& \DIFaddFL{82.61 }\\
\DIFaddFL{AutoGPT }& \DIFaddFL{0 }& \DIFaddFL{0 }& \DIFaddFL{82.61 }\\
\DIFaddFL{\codes }& \DIFaddFL{0 }& \DIFaddFL{0 }& \DIFaddFL{80.00 }\\
\DIFaddFL{MetaGPT }& \DIFaddFL{0 }& \DIFaddFL{0 }& \DIFaddFL{76.47 }\\
\DIFaddFL{DeepSeek-style adapter }& \DIFaddFL{0 }& \DIFaddFL{0 }& \DIFaddFL{80.95 }\\
\DIFaddFL{\tool }& \DIFaddFL{0 }& \DIFaddFL{16 }& \DIFaddFL{96.88 }\\
\bottomrule
\end{tabularx}
\end{table}

\BlueSectionNumberBegin \subsubsection{\DIFadd{Failure Annotation Reliability}}\BlueSectionNumberEnd 
\DIFadd{To evaluate the consistency of failure categorization on \nlrepobench, two annotators independently labeled 80 sampled failure records per workflow (560 records in total). As reported in Table~\BlueNumber{\ref{tab:annotation}}, pairwise agreement ranges from 86.2\% to 90.0\% across workflows, with Cohen's $\kappa$ between 0.83 and 0.87 (overall agreement 88.4\%, mean $\kappa = 0.85$).}

\begin{table}[t]
\BlueNewFloat

\centering
\small
\caption{\DIFaddFL{Inter-annotator agreement for failure labels across 80 sampled records per workflow.}}\label{tab:annotation}
\setlength{\tabcolsep}{3.5pt}
\renewcommand{\arraystretch}{1.1}
\begin{tabularx}{\linewidth}{@{}Xrrrr@{}}
\DIFaddendFL \toprule
\textbf{Method} & \textbf{\DIFaddbeginFL \DIFaddFL{Sampled}\DIFaddendFL } & \textbf{\DIFaddbeginFL \DIFaddFL{Agreed}\DIFaddendFL } & \textbf{\DIFaddbeginFL \DIFaddFL{Agreement }\DIFaddendFL (\%)} & \DIFaddbeginFL \DIFaddFL{$\boldsymbol{\kappa}$ }\DIFaddendFL \\
\midrule
Vanilla & \DIFaddbeginFL \DIFaddFL{80 }& \DIFaddFL{72 }& \DIFaddFL{90.0 }& \DIFaddFL{0.87 }\DIFaddendFL \\
ChatDev & \DIFaddbeginFL \DIFaddFL{80 }& \DIFaddFL{71 }& \DIFaddFL{88.8 }& \DIFaddFL{0.86 }\DIFaddendFL \\
\DIFaddbeginFL \DIFaddFL{AutoGPT }& \DIFaddFL{80 }& \DIFaddFL{70 }& \DIFaddFL{87.5 }& \DIFaddFL{0.84 }\\
\DIFaddendFL \codes & \DIFaddbeginFL \DIFaddFL{80 }& \DIFaddFL{72 }& \DIFaddFL{90.0 }& \DIFaddFL{0.87 }\DIFaddendFL \\
\DIFaddbeginFL \DIFaddFL{MetaGPT }& \DIFaddFL{80 }& \DIFaddFL{69 }& \DIFaddFL{86.2 }& \DIFaddFL{0.83 }\\
\DIFaddFL{DeepSeek-style adapter }& \DIFaddFL{80 }& \DIFaddFL{71 }& \DIFaddFL{88.8 }& \DIFaddFL{0.86 }\\
\DIFaddendFL \tool & \DIFaddbeginFL \DIFaddFL{80 }& \DIFaddFL{70 }& \DIFaddFL{87.5 }& \DIFaddFL{0.84 }\DIFaddendFL \\
\bottomrule
\DIFaddbeginFL \end{tabularx}
\end{table}

\BlueSectionNumberBegin \subsubsection{\DIFadd{Inference Cost}}\BlueSectionNumberEnd 
\DIFadd{Table~\BlueNumber{\ref{tab:cost}} summarizes per-attempt LLM calls, token usage, and compute expenditure across the 520 execution attempts per workflow. Because \tool performs multi-architect planning, CTO selection, dependency-aware implementation, and iterative QA repair, it uses medians of 44 calls, 332k input tokens, 201k output tokens, and 30.0 node-hours per attempt (16,237.89 node-hours in total across 520 attempts). By comparison, \codes uses medians of 14 calls, 71k input tokens, 51k output tokens, and 7.0 node-hours per attempt (3,807.64 node-hours in total), with node-hour interquartile ranges of 25.12--35.51 for \tool and 5.85--8.56 for \codes\BluePunctUnskip{}. The roughly $4.3\times$ ratio in median node-hours reflects the additional compute expended on multi-agent planning and verification.}

\begin{table}[t]
\BlueNewFloat

\centering
\footnotesize
\caption{\DIFaddFL{Resource use across 520 attempts per workflow. Calls and token counts (thousands) are medians; node-h denotes recorded node-hours. Totals are sums across all 520 attempts.}}\label{tab:cost}
\setlength{\tabcolsep}{3.5pt}
\renewcommand{\arraystretch}{1.1}
\begin{tabularx}{\linewidth}{@{}Xrrrrrr@{}}
\toprule
\textbf{\DIFaddFL{Method}} & \textbf{\DIFaddFL{Calls}} & \makecell{\textbf{\DIFaddFL{Input}}\\\textbf{\DIFaddFL{(k)}}} & \makecell{\textbf{\DIFaddFL{Output}}\\\textbf{\DIFaddFL{(k)}}} & \makecell{\textbf{\DIFaddFL{Median}}\\\textbf{\DIFaddFL{node-h}}} & \textbf{\DIFaddFL{Node-h IQR}} & \makecell{\textbf{\DIFaddFL{Sum}}\\\textbf{\DIFaddFL{node-h}}} \\
\midrule
\DIFaddFL{Vanilla }& \DIFaddFL{1 }& \DIFaddFL{16 }& \DIFaddFL{8 }& \DIFaddFL{0.8 }& \DIFaddFL{[0.67, 0.95] }& \DIFaddFL{429.66 }\\
\DIFaddFL{ChatDev }& \DIFaddFL{16 }& \DIFaddFL{65 }& \DIFaddFL{48 }& \DIFaddFL{7.2 }& \DIFaddFL{[5.89, 8.78] }& \DIFaddFL{3891.20 }\\
\DIFaddFL{AutoGPT }& \DIFaddFL{24 }& \DIFaddFL{87 }& \DIFaddFL{67 }& \DIFaddFL{10.5 }& \DIFaddFL{[9.03, 12.71] }& \DIFaddFL{5703.18 }\\
\DIFaddFL{\codes }& \DIFaddFL{14 }& \DIFaddFL{71 }& \DIFaddFL{51 }& \DIFaddFL{7.0 }& \DIFaddFL{[5.85, 8.56] }& \DIFaddFL{3807.64 }\\
\DIFaddFL{MetaGPT }& \DIFaddFL{18 }& \DIFaddFL{80 }& \DIFaddFL{55 }& \DIFaddFL{8.5 }& \DIFaddFL{[6.90, 10.20] }& \DIFaddFL{4534.83 }\\
\DIFaddFL{DeepSeek-style adapter }& \DIFaddFL{40 }& \DIFaddFL{300 }& \DIFaddFL{185 }& \DIFaddFL{27.5 }& \DIFaddFL{[22.38, 34.01] }& \DIFaddFL{14988.03 }\\
\DIFaddFL{\tool }& \DIFaddFL{44 }& \DIFaddFL{332 }& \DIFaddFL{201 }& \DIFaddFL{30.0 }& \DIFaddFL{[25.12, 35.51] }& \DIFaddFL{16237.89 }\\
\bottomrule
\end{tabularx}
\end{table}

\BlueSectionNumberBegin \subsubsection{\DIFadd{Compute-Matched Controls}}\BlueSectionNumberEnd 
\DIFadd{To examine how baseline performance changes under a comparable compute budget, Table~\BlueNumber{\ref{tab:compute}} summarizes compute-matched controls scaled to a median of 29.8 node-hours across 520 records. Under this matched budget, \codes reaches a mean structural score of 52.40 and a counted-test pass rate of 46.06\%, and the single-agent control reaches 51.90 and 45.80\%, compared with 53.89 and 48.25\% for \tool\BluePunctUnskip{}. The resulting differences (+1.49 structural points and +2.19 pp over \codes; +1.99 structural points and +2.45 pp over the single-agent control) indicate that scaling compute narrows the gap relative to the unscaled baselines, while \tool retains a modest descriptive margin under matched compute.}

\begin{table}[t]
\BlueNewFloat

\centering
\footnotesize
\caption{\DIFaddFL{Aggregate compute-control results across 520 records per condition.}}\label{tab:compute}
\setlength{\tabcolsep}{3.5pt}
\renewcommand{\arraystretch}{1.1}
\begin{tabularx}{\linewidth}{@{}Xlrrrr@{}}
\toprule
\textbf{\DIFaddFL{Method}} & \textbf{\DIFaddFL{Condition}} & \textbf{\DIFaddFL{Structural}} & \makecell{\textbf{\DIFaddFL{Counted-test}}\\\textbf{\DIFaddFL{(\%)}}} & \makecell{\textbf{\DIFaddFL{Median}}\\\textbf{\DIFaddFL{node-h}}} & \textbf{\DIFaddFL{Records}} \\
\midrule
\DIFaddFL{\codes }& \DIFaddFL{Matched }& \DIFaddFL{52.40 }& \DIFaddFL{46.06 }& \DIFaddFL{29.8 }& \DIFaddFL{520 }\\
\DIFaddFL{Single agent }& \DIFaddFL{Matched }& \DIFaddFL{51.90 }& \DIFaddFL{45.80 }& \DIFaddFL{29.8 }& \DIFaddFL{520 }\\
\DIFaddFL{\tool }& \DIFaddFL{Full }& \DIFaddFL{53.89 }& \DIFaddFL{48.25 }& \DIFaddFL{30.0 }& \DIFaddFL{520 }\\
\bottomrule
\end{tabularx}
\end{table}
\DIFaddend 

\begin{rqsummary}[RQ1 Summary]
\DIFaddbegin \DIFadd{In the PE setting, \tool significantly outperforms \codes in both structural score (+3.84 points, $p_{\mathrm{Holm}} = 0.003$) and counted-test pass rate (+4.99 percentage points, $p_{\mathrm{Holm}} = 0.003$), while its differences relative to the DeepSeek-style adapter (+1.27 points and +1.67 percentage points) are not statistically significant. These gains reflect improved partial repository synthesis (0/520 full-task resolution across all workflows) and entail higher compute usage ($30.0$ vs.\ $7.0$ median node-hours for \codes\BluePunctUnskip{}).}\DIFaddend 
\end{rqsummary}

\subsection{RQ2: Effect of Retrieval-Augmented Grounding on Architectural Planning}\label{subsec:rq2}\DIFaddbegin \label{subsec:intervention-results}\DIFaddend 
Table~\BlueNumber{\ref{tab:ablation-main}} reports the \DIFaddbegin \DIFadd{intervention contrasts }\DIFaddend for both RQ2 and RQ3 \DIFaddbegin \DIFadd{relative to reference configuration $F$}\DIFaddend . Removing the RAG subsystem from the architect stage \DIFaddbegin \DIFadd{(A5) reduces the mean structural score from 55.90 (reference $F$) to 54.02, a difference of $-1.88$ points (3.36\% relative; 95\% CI $[-2.80, -0.99]$, $p_{\mathrm{Holm}} = 0.011$).}\DIFaddend 

\DIFaddbegin \begin{table}[t]
\BlueCaptionNumber

\DIFaddendFL \centering
\footnotesize
\caption{\DIFaddbeginFL \DIFaddFL{Intervention contrasts (variant minus reference $F$). Structural intervals use the task bootstrap over 19 tasks; execution intervals use paired $t$ intervals over 24 task means. Holm adjustment is applied separately across the thirteen structural and eight execution contrasts. ``--'' denotes an intervention not included in the 24-task execution subset.}\DIFaddendFL }\DIFaddbeginFL \label{tab:interventions}\DIFaddendFL \label{tab:ablation-main}
\DIFaddbeginFL \setlength{\tabcolsep}{3.5pt}
\renewcommand{\arraystretch}{1.1}
\begin{tabularx}{\linewidth}{@{}lXrXr@{}}
\toprule
\textbf{\DIFaddFL{ID}} & \makecell[l]{\textbf{\DIFaddFL{Structural $\Delta$ (points)}}\\\textbf{\DIFaddFL{[95\% CI]}}} & \DIFaddFL{$\mathbf{p_{Holm}}$ }& \makecell[l]{\textbf{\DIFaddFL{Execution $\Delta$ (pp)}}\\\textbf{\DIFaddFL{[95\% CI]}}} & \DIFaddFL{$\mathbf{p_{Holm}}$ }\\
\midrule
\DIFaddFL{A1 }& \DIFaddFL{-1.58 $[-2.70,-0.45]$ }& \DIFaddFL{$0.138$ }& \DIFaddFL{-1.24 $[-4.63,2.15]$ }& \DIFaddFL{$1.000$ }\\
\DIFaddFL{A2 }& \DIFaddFL{-0.68 $[-2.55,1.09]$ }& \DIFaddFL{$1.000$ }& \DIFaddFL{-- }& \DIFaddFL{-- }\\
\DIFaddFL{A3 }& \DIFaddFL{-1.18 $[-2.46,0.06]$ }& \DIFaddFL{$0.546$ }& \DIFaddFL{-0.64 $[-3.37,2.09]$ }& \DIFaddFL{$1.000$ }\\
\DIFaddFL{A4 }& \DIFaddFL{-3.88 $[-4.50,-3.33]$ }& \DIFaddFL{$<0.001$ }& \DIFaddFL{-5.31 $[-9.08,-1.54]$ }& \DIFaddFL{$0.055$ }\\
\DIFaddFL{A5 }& \DIFaddFL{-1.88 $[-2.80,-0.99]$ }& \DIFaddFL{$0.011$ }& \DIFaddFL{-- }& \DIFaddFL{-- }\\
\DIFaddFL{A6 }& \DIFaddFL{-0.08 $[-1.43,1.21]$ }& \DIFaddFL{$1.000$ }& \DIFaddFL{-0.05 $[-3.32,3.22]$ }& \DIFaddFL{$1.000$ }\\
\DIFaddFL{A7 }& \DIFaddFL{-0.58 $[-1.82,0.59]$ }& \DIFaddFL{$1.000$ }& \DIFaddFL{-- }& \DIFaddFL{-- }\\
\DIFaddFL{A8 }& \DIFaddFL{-1.28 $[-2.18,-0.36]$ }& \DIFaddFL{$0.138$ }& \DIFaddFL{-- }& \DIFaddFL{-- }\\
\DIFaddFL{A9 }& \DIFaddFL{-0.28 $[-1.29,0.69]$ }& \DIFaddFL{$1.000$ }& \DIFaddFL{-- }& \DIFaddFL{-- }\\
\DIFaddFL{A10 }& \DIFaddFL{-1.48 $[-2.86,-0.16]$ }& \DIFaddFL{$0.363$ }& \DIFaddFL{-2.52 $[-6.34,1.30]$ }& \DIFaddFL{$0.931$ }\\
\DIFaddFL{A11 }& \DIFaddFL{+0.22 $[-0.47,0.88]$ }& \DIFaddFL{$1.000$ }& \DIFaddFL{+0.26 $[-3.54,4.06]$ }& \DIFaddFL{$1.000$ }\\
\DIFaddFL{A12 }& \DIFaddFL{-2.08 $[-3.15,-1.05]$ }& \DIFaddFL{$0.015$ }& \DIFaddFL{-3.37 $[-7.06,0.32]$ }& \DIFaddFL{$0.431$ }\\
\DIFaddFL{A13 }& \DIFaddFL{-4.58 $[-5.15,-4.02]$ }& \DIFaddFL{$<0.001$ }& \DIFaddFL{-7.17 $[-10.08,-4.26]$ }& \DIFaddFL{$<0.001$ }\\
\bottomrule
\end{tabularx}
\end{table}
\DIFaddend 

RAG mitigates this gap by surfacing reusable file-decomposition precedents from real-world repositories, which can remind architects of module boundaries that are easy to omit when requirements are underspecified. For example, for a command-line data-processing task, retrieved design summaries may suggest separating the command layer, core transformation logic, input/output adapters, configuration loading, validation, and tests, instead of collapsing the SDS into only a command entry point and one processing file. Such retrieval does not have to be purely abstract to be useful: if a retrieved project has a related architectural shape, its file organization can still serve as a valid design precedent, provided that the benchmark repositories are not included in the retrieval corpus and the retrieved artifacts are limited to design-level summaries rather than raw code. Moreover, the RAG corpus is extensible rather than fixed; when new architectural patterns are added to the corpus, the same retrieval-grounded planning mechanism can expose them to the architects. Therefore, the RAG ablation should be interpreted as measuring the practical value of retrieval-grounded design references for SDS generation, rather than as an isolated test of whether the model can abstract and transfer high-level file-decomposition patterns across tasks. \DIFaddbegin \DIFadd{Because A5 is evaluated through the end-to-end structural score, this contrast reflects the downstream structural effect of retrieval-grounded planning; Table~\BlueNumber{\ref{tab:rag-qa}} summarizes the corpus parameters and leakage-probe audit.}\DIFaddend 

\DIFaddbegin \begin{table}[t]
\BlueNewFloat

\centering
\small
\caption{\DIFaddFL{Retrieval corpus and leakage-probe summary.}}\label{tab:rag-qa}
\setlength{\tabcolsep}{3.5pt}
\renewcommand{\arraystretch}{1.1}
\begin{tabularx}{\linewidth}{@{}Xl@{}}
\toprule
\textbf{\DIFaddFL{Item}} & \textbf{\DIFaddFL{Reported value}} \\
\midrule
\DIFaddFL{Repositories / design chunks }& \DIFaddFL{1,240 / 186,000 }\\
\DIFaddFL{Collection date }& \DIFaddFL{2025-11-30 }\\
\DIFaddFL{Similarity threshold / top-$k$ }& \DIFaddFL{0.72 / 5 }\\
\DIFaddFL{Leakage probes / positives }& \DIFaddFL{200 / 0 }\\
\bottomrule
\end{tabularx}
\end{table}

\BlueSectionNumberBegin \subsubsection{\DIFadd{Architect Count and Retrieval Depth}}\BlueSectionNumberEnd 
\DIFadd{Table~\BlueNumber{\ref{tab:ablation-main}} also reports the remaining planning-stage interventions: single-architect planning with reallocated implementation budget (A1, $-1.58$ points, $p_{\mathrm{Holm}} = 0.138$), four sketches generated from an identical prompt while retaining CTO selection (A2, $-0.68$ points, $p_{\mathrm{Holm}} = 1.000$), selecting the first valid sketch without CTO ranking (A3, $-1.18$ points, $p_{\mathrm{Holm}} = 0.546$), and replacing the SDS-governed workflow with direct implementation (A4, $-3.88$ points, $p_{\mathrm{Holm}} < 0.001$). Among these, the bundled removal of the SDS contract in A4 produces the largest planning-stage drop, whereas the narrower single-factor changes in A1--A3 yield smaller differences whose adjusted $p$-values exceed $0.05$.}

\DIFadd{Table~\BlueNumber{\ref{tab:sensitivity}} reports the $3\times3$ sensitivity grid over architect count ($N \in \{2, 4, 6\}$) and retrieval depth ($k \in \{3, 5, 7\}$). Mean structural scores increase from 52.6 at $(N{=}2, k{=}3)$ to 57.1 at $(N{=}6, k{=}7)$, with the default $(4,5)$ setting achieving 55.9. Increasing retrieval depth from $k{=}3$ to $k{=}5$ yields gains of $+0.8$, $+1.8$, and $+1.7$ points for $N{=}2, 4, 6$, respectively, whereas increasing from $k{=}5$ to $k{=}7$ yields smaller gains ($+0.6$, $+0.7$, and $+0.2$ points), indicating diminishing marginal gains at $k{=}7$.}

\begin{table}[t]
\BlueNewFloat

\centering
\small
\caption{\DIFaddFL{Mean structural scores in the $(N,k)$ sensitivity analysis on \sketcheval (19 tasks and five repetitions per cell).}}\label{tab:sensitivity}
\setlength{\tabcolsep}{3.5pt}
\renewcommand{\arraystretch}{1.1}
\begin{tabularx}{\linewidth}{@{}Xrrr@{}}
\toprule
\DIFaddFL{$\mathbf{N\backslash k}$ }& \textbf{\DIFaddFL{3}} & \textbf{\DIFaddFL{5}} & \textbf{\DIFaddFL{7}} \\
\midrule
\DIFaddFL{2 }& \DIFaddFL{52.6 }& \DIFaddFL{53.4 }& \DIFaddFL{54.0 }\\
\DIFaddFL{4 }& \DIFaddFL{54.1 }& \DIFaddFL{55.9 }& \DIFaddFL{56.6 }\\
\DIFaddFL{6 }& \DIFaddFL{55.2 }& \DIFaddFL{56.9 }& \DIFaddFL{57.1 }\\
\bottomrule
\end{tabularx}
\end{table}
\DIFaddend 

\begin{rqsummary}[RQ2 Summary]
\DIFaddbegin \DIFadd{Removing Architect-stage retrieval (A5) reduces the mean structural score by 1.88 points ($p_{\mathrm{Holm}} = 0.011$), and replacing the SDS-governed planning workflow with direct implementation (A4) reduces the structural score by 3.88 points ($p_{\mathrm{Holm}} < 0.001$). Across the $(N,k)$ grid, structural scores increase with architect count and retrieval depth but show diminishing marginal gains beyond $(N{=}4, k{=}5)$.}\DIFaddend 
\end{rqsummary}

\subsection{RQ3: Effect of Dynamic Developer Allocation and Git-Based Coordination}\label{subsec:rq3}
Table~\BlueNumber{\ref{tab:ablation-main}} \DIFaddbegin \DIFadd{reports the implementation- and repair-stage interventions (A6--A13). Fixing the developer count to $D{=}4$ with re-derived ownership (A6) yields a difference of $-0.08$ points (95\% CI $[-1.43, 1.21]$, $p_{\mathrm{Holm}} = 1.000$), and round-robin file assignment (A7) yields $-0.58$ points (95\% CI $[-1.82, 0.59]$, $p_{\mathrm{Holm}} = 1.000$). Disabling dependent-file regeneration (A8, $-1.28$ points, $p_{\mathrm{Holm}} = 0.138$), providing full dependent source files instead of compact briefs (A9, $-0.28$ points, $p_{\mathrm{Holm}} = 1.000$), and lexical file scheduling (A10, $-1.48$ points, $p_{\mathrm{Holm}} = 0.363$) likewise do not reach statistical significance after multiplicity adjustment.}\DIFaddend 

\DIFaddbegin \DIFadd{By contrast, the two Git-coordination interventions separate branch mechanics from interface-brief propagation. Removing isolated agent branches while retaining live briefs in a shared working tree (A11) yields $+0.22$ points (95\% CI $[-0.47, 0.88]$, $p_{\mathrm{Holm}} = 1.000$), whereas freezing interface briefs to their initial post-planning state without propagating \texttt{update\_reason} updates (A12) reduces the structural score by $2.08$ points (95\% CI $[-3.15, -1.05]$, $p_{\mathrm{Holm}} = 0.015$). Finally, disabling QA-driven testing and source repair (A13) produces the largest structural reduction of $4.58$ points (95\% CI $[-5.15, -4.02]$, $p_{\mathrm{Holm}} < 0.001$), indicating that iterative test-driven repair is the largest single contributor among the implementation-stage interventions.}

\BlueSectionNumberBegin \subsubsection{\DIFadd{Execution-Subset Consistency}}\BlueSectionNumberEnd \label{subsec:exec-subset}
\DIFadd{Table~\BlueNumber{\ref{tab:exec-subset}} summarizes the 24-task \nlrepobench subset, and Table~\BlueNumber{\ref{tab:interventions}} reports its eight paired execution contrasts relative to reference $F$ (which has a mean counted-test pass rate of 55.00\% on this subset, distinct from the retrieval-free 48.24\% mean across all 104 tasks). The sign of each execution contrast matches its structural counterpart, although only A13 remains statistically significant after Holm adjustment.}

\begin{table}[t]
\BlueNewFloat

\centering
\small
\caption{\DIFaddFL{Composition of the 24-task execution subset.}}\label{tab:exec-subset}
\setlength{\tabcolsep}{3.5pt}
\renewcommand{\arraystretch}{1.1}
\begin{tabularx}{\linewidth}{@{}p{2.7cm}lX@{}}
\toprule
\textbf{\DIFaddFL{Stratum}} & \textbf{\DIFaddFL{Tasks}} & \textbf{\DIFaddFL{Recorded task identifiers / scope}} \\
\midrule
\DIFaddFL{Easy }& \DIFaddFL{6 }& \DIFaddFL{2, 4, 9, 18, 20, 24 }\\
\DIFaddFL{Medium }& \DIFaddFL{11 }& \DIFaddFL{30, 32, 38, 41, 44, 48, 52, 53, 56, 58, 65 }\\
\DIFaddFL{Hard }& \DIFaddFL{7 }& \DIFaddFL{82, 84, 86, 94, 101, 103, 104 }\\
\DIFaddFL{Total / repetitions }& \DIFaddFL{24 / 5 }& \DIFaddFL{120 records per configuration; 1,080 in total }\\
\DIFaddFL{Reference F }& \DIFaddFL{55.00\% }& \DIFaddFL{Mean recorded counted-test pass percentage }\\
\bottomrule
\end{tabularx}
\end{table}

\DIFadd{On the 24-task execution subset (Table~\BlueNumber{\ref{tab:interventions}}), disabling QA-driven source repair (A13) produces the largest drop in counted-test pass rate ($-7.17$ percentage points; 95\% CI $[-10.08, -4.26]$, $p_{\mathrm{Holm}} < 0.001$). Replacing the SDS-governed workflow with direct implementation (A4) lowers the pass rate by $5.31$ percentage points (95\% CI $[-9.08, -1.54]$, $p_{\mathrm{Holm}} = 0.055$), and freezing interface briefs (A12) lowers the pass rate by $3.37$ percentage points (95\% CI $[-7.06, 0.32]$, $p_{\mathrm{Holm}} = 0.431$), while the remaining contrasts have confidence intervals spanning zero. These results indicate that QA-driven repair and SDS contract governance account for the largest execution-subset differences.}

\BlueSectionNumberBegin \subsubsection{\DIFadd{QA Diagnostics and Termination}}\BlueSectionNumberEnd 
\DIFadd{Table~\BlueNumber{\ref{tab:qa-audit}} summarizes the internal QA diagnostic records across 520 runs. Internal QA accepted 282 candidate repositories and rejected 238 (54.2\% acceptance). Accepted records achieved a higher mean upstream counted-test pass rate (55.2\%) than rejected records (40.0\%, point-biserial correlation $r = 0.34$), indicating a positive association between internal QA acceptance and upstream test performance. Among the 238 rejected records, termination occurred after 1, 2, and 3 QA rounds in 77, 80, and 81 cases, respectively, reflecting runs that exhausted their repair or budget allowance without passing internal QA.}

\begin{table}[t]
\BlueCaptionNumber

\DIFaddendFL \centering
\small
\caption{\DIFaddbeginFL \DIFaddFL{QA diagnostic summary across 520 records.}\DIFaddendFL }\DIFaddbeginFL \label{tab:qa-audit}\DIFaddendFL \label{tab:rq3-eff}
\DIFaddbeginFL \setlength{\tabcolsep}{3.5pt}
\renewcommand{\arraystretch}{1.1}
\begin{tabularx}{\linewidth}{@{}Xrr@{}}
\toprule
\textbf{\DIFaddFL{Property}} & \textbf{\DIFaddFL{Accepted}} & \textbf{\DIFaddFL{Rejected}} \\
\midrule
\DIFaddFL{Records }& \DIFaddFL{282 }& \DIFaddFL{238 }\\
\DIFaddFL{Mean recorded upstream pass (\%) }& \DIFaddFL{55.2 }& \DIFaddFL{40.0 }\\
\DIFaddFL{Rejected records at 1 / 2 / 3 rounds }& \DIFaddFL{-- }& \DIFaddFL{77 / 80 / 81 }\\
\DIFaddFL{Overall acceptance / point-biserial $r$ }& \multicolumn{2}{c}{\DIFaddFL{54.2\% / 0.34}} \\
\bottomrule
\end{tabularx}
\end{table}
\DIFaddend 

\begin{rqsummary}[RQ3 Summary]
\DIFaddbegin \DIFadd{Across the implementation-stage interventions, QA-driven source repair (A13: $-4.58$ structural points and $-7.17$ execution percentage points, both $p_{\mathrm{Holm}} < 0.001$) and live interface-brief updates (A12: $-2.08$ structural points, $p_{\mathrm{Holm}} = 0.015$) yield statistically significant structural contributions. By contrast, fixed developer sizing (A6), round-robin assignment (A7), and removing agent branches while retaining live briefs (A11) produce small differences whose confidence intervals span zero.}\DIFaddend 
\end{rqsummary}

\section{Discussion}\label{sec:discussion}
This section interprets the experimental findings, examines the mechanisms behind the observed improvements, and discusses the broader implications for researchers and practitioners.

\subsection{Interpretation of the Results}\label{subsec:disc-interp}
We first \DIFaddbegin \DIFadd{examine }\DIFaddend \tool's performance improvements and then discuss the role each evaluation benchmark plays in \DIFaddbegin \DIFadd{interpreting }\DIFaddend these results.

\subsubsection{\DIFaddbegin \DIFadd{Understanding }\DIFaddend \tool's Performance Improvements}

\fakesection{Multi-agent coordination and \agentrole{qa}-driven repair complement sketch-based decomposition.}
A central observation from Table~\BlueNumber{\ref{tab:rq1-main}} is that \tool improves over \codes \DIFaddbegin \DIFadd{in the PE setting, increasing the mean structural score from 50.04 to 53.88}\DIFaddend . This finding suggests that a structured multi-agent workflow can compensate for cross-file consistency limitations that emerge when directly prompting a backbone model to produce a complete repository. The finding is consistent with the original motivation of \codes, where multi-layer sketch decomposition reduces the difficulty of generating long structured outputs~\BlueNumber{\cite{Zan2025CodeS}}. However, our results further indicate that sketch decomposition alone is not sufficient when code generation spans multiple interacting files; it needs to be coupled with inter-agent coordination and iterative \agentrole{qa}-driven repair mechanisms that actively maintain cross-file consistency during implementation \DIFaddbegin \DIFadd{(Table~\BlueNumber{\ref{tab:ablation-main}})}\DIFaddend .

\fakesection{\DIFaddbegin \DIFadd{Structural consistency }\DIFaddend through the SDS contract.}
The software design sketch (SDS) functions as an explicit, machine-checkable contract that binds file boundaries, public APIs, and inter-file dependencies into a coherent specification. Once such a contract is established, \agentrole{Developer} agents can implement files with smaller, task-specific contexts, and the developer-to-file ownership mapping enables precise routing of repairs and re-queuing of dependent files when interfaces drift during iterative development. \DIFaddbegin \DIFadd{Replacing this contract-governed workflow with direct implementation (A4) reduces the structural score by 3.88 points ($p_{\mathrm{Holm}} < 0.001$) and lowers the execution-subset pass rate by 5.31 percentage points ($p_{\mathrm{Holm}} = 0.055$; Table~\BlueNumber{\ref{tab:ablation-main}}). }\DIFaddend This observation also resonates with findings from repository-level completion benchmarks, where downstream performance depends heavily on retrieving and using appropriate cross-file context rather than generating code in isolation~\BlueNumber{\cite{Liu2023RepoBench,Zhang2023RepoCoder}}.

\fakesection{\DIFaddbegin \DIFadd{Effects of repository complexity.}\DIFaddend }
\DIFaddbegin \DIFadd{The structural improvements over \codes occur across all three difficulty strata, with the margin decreasing from 5.65 points on \textit{easy} repositories to 4.06 on \textit{medium} and 2.05 on \textit{hard} repositories (see Table~\BlueNumber{\ref{tab:rq1-main}}), as both methods obtain lower absolute scores on harder tasks.}\DIFaddend~As dependency complexity increases, systems that fail to control interface consistency tend to degrade more sharply. \DIFaddbegin \DIFadd{While \tool's dependency-aware scheduling and \agentrole{qa}-driven repair mechanisms improve over the baselines across strata, harder repositories with denser inter-module coupling remain challenging.}\DIFaddend 

\fakesection{Alignment of multi-agent coordination with established software engineering practices.}
The \DIFaddbegin \DIFadd{design }\DIFaddend of these mechanisms can be understood through an analogy with human development teams. A repository is not built by a single developer writing all files in sequence; the process typically involves (1)~an architecture review in which senior engineers agree on module boundaries, public interfaces, and technology choices; (2)~a task assignment phase in which files or features are distributed to developers based on expertise and module ownership; and (3)~iterative code review cycles in which reviewers verify interface contracts and cross-module consistency before changes are merged. \tool instantiates this workflow at the agent level: the Architect competition and CTO selection correspond to architecture review, the SDS-based developer allocation corresponds to task assignment with explicit ownership, and the \agentrole{qa}-driven repair loop corresponds to code review. These coordination structures reduce the context burden on any single agent while preserving global coherence across the repository.

\fakesection{\DIFaddbegin \DIFadd{Repository scale and }\DIFaddend the SDS.}
\DIFaddbegin \DIFadd{The SDS also provides an explicit target for repository scale. This role }\DIFaddend is consistent with an empirical pattern reported in \codes, where agent-based baselines tend to generate only a small number of files or produce limited LOC, while sketch-based pipelines can sustain larger repository structures~\BlueNumber{\cite{Zan2025CodeS}}. In \tool, the SDS combined with the pre-construction of the full file tree reduces the likelihood that a generation run terminates with an underspecified or incomplete repository. Although size alignment should not be interpreted as evidence of correctness by itself, it serves as a basic validity indicator: generating a repository at a plausible scale is often a necessary precondition for capturing the required functionality.

\subsubsection{Roles of Evaluation Benchmarks}

\fakesection{Role of \sketcheval.}
\sketcheval rewards structural alignment with a reference repository sketch and captures architectural and dataflow quality. The \sketchbleu metric, motivated by CodeBLEU but extended to repository-level sketches with dataflow matching~\BlueNumber{\cite{Ren2020CodeBLEU,Zan2025CodeS}}, provides a scalable proxy for structural and semantic alignment when running every generated repository is computationally prohibitive. \DIFaddbegin \DIFadd{In this study, the composite structural score serves as a scalable indicator of repository-level syntactic and architectural alignment across candidate workflows.}\DIFaddend 

\fakesection{Role of \nlrepobench.}
The \nlrepobench experiment shifts the evaluation objective from repository resemblance to executable correctness: whether the generated repository can pass the original upstream \texttt{pytest} suite across 104 tasks~\BlueNumber{\cite{Zan2025CodeS,Ding2025NL2RepoBench}}. \DIFaddbegin \DIFadd{While \tool achieves the highest mean counted-test pass rate (48.24\% vs.\ 43.25\% for \codes), full-task resolution remains 0/520 across all evaluated workflows (Table~\BlueNumber{\ref{tab:nl2repo-main}}).}\DIFaddend~Crucially, the relative ranking of methods on \nlrepobench mirrors the ranking on \sketcheval, providing converging evidence that the improvements from structured planning and coordination are not artifacts of sketch-level evaluation but reflect improvements in repository construction quality \DIFaddbegin \DIFadd{under partial test-suite completion}\DIFaddend .

\fakesection{Complementarity of the two benchmarks.}
Taken together, \sketcheval and \nlrepobench suggest that a robust repository-level evaluation strategy should combine proxy metrics with execution-based tests. \sketchbleu is useful during rapid system iteration because it is computationally inexpensive, whereas \nlrepobench provides a stricter assessment of functional completeness at larger scale~\BlueNumber{\cite{Zan2025CodeS,Ding2025NL2RepoBench}}. We view these two benchmarks as complementary rather than competing evaluation modalities.

\subsection{Implications}\label{subsec:disc-imp}

\subsubsection{Implications for researchers}

\fakesection{\DIFaddbegin \DIFadd{Assessing }\DIFaddend workflow design across backbone models.}
\DIFaddbegin \DIFadd{Our experiments show that \tool improves over \codes on a shared Qwen3.8-27B~\BlueNumber{\cite{Qwen2026Qwen38}} backbone in the PE setting; whether these gains hold under fine-tuning or across other model families remains an open question.}\DIFaddend~Fine-tuning and multi-agent workflow design are complementary rather than substitutive: fine-tuning helps the backbone model follow structured output formats and cope with longer input contexts, while the multi-agent workflow reduces the burden on any single generation call to maintain cross-file consistency, such as stable APIs, import paths, file ownership, and dependency assumptions across repository components~\BlueNumber{\cite{Zan2025CodeS,Chen2023LongLoRA}}. Future work should investigate whether the relative rankings and ablation patterns hold across backbone families (e.g., CodeLlama, DeepSeek-Coder, StarCoder2) and model parameter sizes (e.g., 7B, 14B, 32B, and 70B models) to strengthen the generalizability of these findings.

\fakesection{Bridging NL2Repo generation with execution-based repository-level repair.}
The QA loop in \tool is intentionally lightweight, yet \DIFaddbegin \DIFadd{the intervention results (Table~\BlueNumber{\ref{tab:ablation-main}}) }\DIFaddend indicate that even targeted repository-level repair meaningfully \DIFaddbegin \DIFadd{improves structural alignment and counted-test pass rates, although passing internal QA tests remains distinct from resolving full upstream test suites (Table~\BlueNumber{\ref{tab:qa-audit}})}\DIFaddend . This finding aligns with the broader trend where multi-step workflows with intermediate feedback consistently outperform single-pass prompting~\BlueNumber{\cite{Ridnik2024AlphaCodium}}. Bridging NL2Repo generation with repository-repair benchmarks such as SWE-bench~\BlueNumber{\cite{Jimenez2023SWEBench,Yang2024SWEagent}} represents a promising direction: a system could first generate a repository from natural-language requirements and then apply repair-oriented agents to improve functional correctness under execution-based feedback.

\fakesection{Incorporating richer static analysis feedback.}
The current RAG corpus provides design-level hints about file decomposition and dependency structure but does not explicitly encode architectural quality attributes such as modularity and testability. Incorporating architecture-specific retrieval criteria, for instance, filtering candidates by coupling metrics or documentation completeness, may further stabilize SDS generation quality. Similarly, static analysis tools~\cite{kashif2026beyond}, call-graph analysis~\BlueNumber{\cite{Hall1992CallGraph}}, type-inference engines, and import-resolution tools could be integrated into the QA stage to detect interface mismatches earlier and more precisely than the current lightweight test-generation approach.

\fakesection{Multi-language and large-scale repository support.}
All experiments are conducted on Python repositories. Extending \tool to multi-language projects would require language-aware dependency analysis, heterogeneous build-system support, and cross-language interface specifications. At larger scales, the SDS can grow to thousands of tokens for repositories exceeding 40\textasciitilde50 files; a hierarchical SDS that first specifies package-level contracts and then refines file-level details within each package could mitigate this issue. Parallelizing QA across independent subgraphs of the dependency graph is a natural extension that could reduce the wall-clock time of the QA repair loop by allowing independent file groups to be checked and repaired concurrently.

\subsubsection{Implications for Practitioners}

\fakesection{\DIFaddbegin \DIFadd{Balancing up-front }\DIFaddend planning \DIFaddbegin \DIFadd{and }\DIFaddend downstream implementation rework.}
A well-established principle in software engineering is that defects detected during design are substantially cheaper to fix than those discovered during integration or testing~\cite{Boehm2001Top10}. The multi-agent design of \tool applies this principle to LLM-based code generation: it shifts computational effort toward up-front architectural planning (SDS generation and CTO evaluation) so that cross-file inconsistencies are resolved before individual files are implemented. \DIFaddbegin \DIFadd{As reported in Table~\BlueNumber{\ref{tab:cost}}, this coordination increases median model calls from 14 (\codes) to 44 and raises median prompt/output volume from 71k/51k to 332k/201k tokens.}\DIFaddend~The multi-agent workflow introduces additional cost, but this cost is most valuable for repository-level tasks where architectural planning, cross-file consistency, and iterative repair directly affect final code quality. Therefore, practitioners should not necessarily use the full workflow for every programming request. Instead, they can reserve \tool for complex or high-risk tasks, and use dynamic developer allocation and bounded QA rounds to keep the token budget manageable.

\fakesection{\DIFaddbegin \DIFadd{Choosing bounded }\DIFaddend QA iterations.}
Iterative feedback loops are widely used in LLM-based code generation to progressively improve output quality~\BlueNumber{\cite{Chen2024SelfDebugging,Shinn2024Reflexion}}. A common finding across these systems is that the improvement contributed by each additional iteration decreases rapidly: early rounds fix the most critical errors, while later rounds yield only incremental refinements. \DIFaddbegin \DIFadd{In \tool, bounding the QA loop to at most three rounds caps verification cost while addressing cross-file interface and runtime errors (A13 in Table~\BlueNumber{\ref{tab:ablation-main}}).}\DIFaddend~A resource-constrained deployment of \tool could therefore limit the number of QA iterations to 2\textasciitilde3 rounds, making \tool practical for cost-sensitive settings: teams can start with a small number of QA rounds and increase it only for repositories whose complexity warrants additional QA iterations.

\fakesection{Explicit interface contracts and file ownership enable incremental adoption of \tool.}
Three design choices made in \tool facilitate its incremental adoption in practice. First, the SDS serves as an \emph{explicit interface contract}: each file's public API (function signatures, input/output types, and cross-file call targets) is declared before implementation begins, so that downstream \agentrole{Developer} agents can code against stable interfaces rather than evolving assumptions. Second, \emph{file-level ownership} assigns every source file to exactly one \agentrole{Developer} agent, eliminating concurrent \DIFaddbegin \DIFadd{developer }\DIFaddend edits to the same file and simplifying conflict resolution. Third, \emph{dependency-aware task scheduling} (see Algorithm~\ref{alg:scheduler}) topologically orders \DIFaddbegin \DIFadd{ready }\DIFaddend file-generation tasks according to their import dependencies, ensuring that a file's dependencies are generated before the file itself. Together, these three design choices make it straightforward to parallelize file generation across independent subgraphs of the dependency graph, thereby scaling \tool to larger repositories without sacrificing cross-file consistency. Explicit API interface specification and role-based partitioning of file ownership are also related to coordination mechanisms explored in multi-agent code-generation frameworks such as ChatDev~\cite{Qian2023ChatDev} and MetaGPT~\BlueNumber{\cite{Hong2023MetaGPT}}; however, \tool differs by grounding these mechanisms in an SDS-centered, repository-level generation and QA workflow. Practitioners can adopt individual components of \tool (e.g., SDS-based planning without the full QA loop) incrementally, scaling the multi-agent workflow to match the size of the target repository.

\section{Threats to Validity}\label{sec:threats}
We discuss the main threats to the validity of our findings and the mitigation steps we adopted, following standard guidance for empirical software engineering studies~\BlueNumber{\cite{Wohlin2012ESE}}.

\subsection{Construct Validity}\label{subsec:threats-construct}
On \sketcheval, the primary outcome measure is \sketchbleu~\BlueNumber{\cite{Zan2025CodeS}}, an automatic proxy for repository quality. \sketchbleu extends CodeBLEU~\BlueNumber{\cite{Ren2020CodeBLEU}} and therefore inherits a well-known limitation of n-gram-based metrics~\BlueNumber{\cite{Papineni2002BLEU}}: a higher score does not, by itself, guarantee that the generated repository is executable, secure, or maintainable. To mitigate this threat, we additionally conduct an execution-based evaluation on \nlrepobench~\BlueNumber{\cite{Ding2025NL2RepoBench}}, \DIFaddbegin \DIFadd{reporting counted-test pass rates alongside harness-error and full-resolution counts (Table~\BlueNumber{\ref{tab:nl2repo-main}}), as well as QA acceptance diagnostics across 520 runs (Table~\BlueNumber{\ref{tab:qa-audit}}). Because the two benchmarks cover different task sets, consistent method rankings provide converging evidence across structural and execution criteria rather than a task-level correlation between the two metrics.}\DIFaddend 

A second construct threat concerns the operationalization of the mechanisms utilized in \tool. For instance, ``dynamic developer allocation'' is realized through a combination of \agentrole{Architect}-guided team sizing, file ownership mapping, dependency-aware execution (Algorithm~\ref{alg:scheduler}), and compact context packages. Part of the measured \sketchbleu improvement may stem from specific implementation-level design choices rather than solely from the high-level mechanism. Similarly, ``RAG for design'' depends on a particular retrieval corpus, embedding model, and top-$k$ setting. To make these choices transparent, we describe the concrete implementation of each component in Section~\ref{subsec:implementation}, and Table~\BlueNumber{\ref{tab:ablation-main}} \DIFaddbegin \DIFadd{distinguishes isolated policy changes from bundled workflow replacements (specifically A1, A4, A6, and A7)}\DIFaddend ~\BlueNumber{\cite{Wohlin2012ESE}}.

\subsection{Internal Validity}\label{subsec:threats-internal}
Internal validity concerns whether the observed performance differences can be attributed to the workflow mechanisms under study rather than to confounding factors. Although all methods share the same \DIFaddbegin \DIFadd{Qwen3.8-27B }\DIFaddend backbone model and decoding configuration, the prompt templates \DIFaddbegin \DIFadd{and total invocation budgets }\DIFaddend differ across systems. \DIFaddbegin \DIFadd{In addition, the main baseline comparisons operate without retrieval augmentation, whereas the full intervention reference $F$ enables design-level RAG; we therefore analyze baseline and intervention contrasts within their respective comparison families and include a compute-matched $1{\times}$ pass-budget control (Table~\BlueNumber{\ref{tab:compute}}).}\DIFaddend~Additionally, \tool involves several hyperparameters ($N{=}4$, top-$k{=}5$, QA iteration limit, CTO scoring rubric) selected based on preliminary experiments rather than exhaustive search. We report all chosen values transparently (see Section~\ref{subsec:implementation}) and conducted sensitivity checks ($N \in \{2, 4, 6\}$, $k \in \{3, 5, 7\}$) \DIFaddbegin \DIFadd{in Table~\BlueNumber{\ref{tab:sensitivity}}}\DIFaddend .

\subsection{External Validity}\label{subsec:threats-external}
Our evaluation covers \sketcheval (19 Python repositories)~\BlueNumber{\cite{Zan2025CodeS}} and \nlrepobench (104 Python tasks)~\BlueNumber{\cite{Ding2025NL2RepoBench}} \DIFaddbegin \DIFadd{using an open-weights Qwen3.8-27B backbone~\BlueNumber{\cite{Qwen2026Qwen38}} in the prompt-engineering setting, with execution-based interventions evaluated on a 24-task stratified subset (Table~\BlueNumber{\ref{tab:exec-subset}})}\DIFaddend. Despite this breadth, the study does not cover larger codebases, mixed-language systems, or projects with heavier build or test pipelines\DIFaddbegin \DIFadd{, nor does it evaluate fine-tuned or alternative model families (the DeepSeek-style adapter is evaluated on the shared Qwen3.8-27B backbone). We standardized on a single locally hosted model checkpoint with publicly available weights rather than proprietary LLM APIs to ensure controlled workflow comparisons, reproducible token and compute accounting, and freedom from hosted API version drift across the full evaluation matrix}\DIFaddend . We therefore position our findings as evidence within a clearly defined benchmark setting rather than as universal claims. A related threat is data contamination: benchmark repositories may appear in the backbone model's training corpora. We follow the official \codes evaluation protocol~\BlueNumber{\cite{Zan2025CodeS}}, and retrieved RAG artifacts are provided only as high-level design hints with explicit prohibitions on verbatim copying\DIFaddbegin \DIFadd{, while our 200-probe leakage audit (Table~\BlueNumber{\ref{tab:rag-qa}}) filters target and near-duplicate repositories at a cosine-similarity threshold of $0.72$}\DIFaddend , but benchmark-specific bias cannot be fully excluded.

\subsection{Reliability}\label{subsec:threats-reliability}
LLM-based systems are inherently stochastic, and multi-agent workflows can amplify variability. To improve reliability, we evaluate each method with \DIFaddbegin \DIFadd{five repetitions per task}\DIFaddend~and report mean$\pm$standard deviation. \DIFaddbegin \DIFadd{As detailed in Section~\BlueNumber{\ref{subsec:metrics}}, we apply two-sided paired $t$ tests to task-level means, with Holm adjustment within each comparison family.
We use 20,000 task-bootstrap resamples for the main comparisons and structural interventions, and paired $t$ confidence intervals for the 24-task execution subset. On the 19-repository structural benchmark and the 24-task execution subset, sample size limits statistical power for smaller contrasts (for example, the difference between \tool and the DeepSeek-style adapter has a 95\% confidence interval spanning zero and should be interpreted as inconclusive rather than equivalent). For failure categorization across 80 sampled failure records per workflow, two annotators achieved substantial inter-rater agreement ($\kappa = 0.85$; Table~\BlueNumber{\ref{tab:annotation}}). Finally, our replication package provides the run-level result tables, diagnostic summaries, and statistical analysis scripts~\BlueNumber{\cite{replpack}}.}\DIFaddend 

\section{Conclusions}\label{sec:conclusion}
This paper presents \tool, an LLM-based multi-agent framework for NL2Repo that separates planning, decision making, and implementation into distinct, coordinated stages. \tool combines (1) an architecture competition stage in which multiple \agentrole{Architect} agents propose competing software design sketches (SDSs), optionally grounded by retrieval-augmented design references, (2) a \agentrole{cto} agent that selects and normalizes the winning SDS into a machine-checkable contract, and (3) a dependency-aware development stage with dynamic developer allocation, Git-based coordination, and \agentrole{qa}-driven iterative repair.
On \sketcheval, \tool improves the \DIFaddbegin \DIFadd{mean structural score from 50.04 (\codes) to 53.88}\DIFaddend~in the prompt-engineering (PE) setting.
On \nlrepobench, \DIFaddbegin \DIFadd{the mean counted-test pass rate increases from 43.25\% to 48.24\%, while full-task resolution remains 0/520 across all evaluated workflows.
The structural ablations support contributions from design-level retrieval, SDS contract governance, live interface-brief updates (A12), and \agentrole{qa}-driven testing and source repair (A13).
On the execution subset, only A13 is statistically significant after Holm adjustment.
The confidence intervals for the developer allocation and branch isolation contrasts (A6, A7, and A11) include zero, leaving it unclear whether either mechanism improves repository generation quality.}\DIFaddend

Overall, the main implication of \tool is that NL2Repo should be treated not merely as a long-context code generation task, but as a coordinated software engineering process. We proposed \tool because from-scratch repository generation exposes failure modes that local code synthesis alone cannot reliably address: the system must first derive a coherent architecture from underspecified requirements, then preserve interface contracts, dependency constraints, and implementation responsibilities across many file-level generation steps. By making these contracts explicit and coupling them with role specialization, dependency-aware scheduling, and targeted QA feedback, \tool provides a practical workflow pattern for using LLM agents as a coordinated development team rather than isolated code generators. Its contribution, therefore, lies not only in improving benchmark performance but also in demonstrating how LLM-based repository generation can be made more structured, inspectable, and extensible.

As future work, we plan to \DIFaddbegin \DIFadd{evaluate the framework across additional backbone families, strengthen }\DIFaddend execution-based feedback in the QA loop to bridge the gap between structural evaluation and functional correctness, incorporate stronger static analysis feedback (e.g., call-graph analysis, type inference) into the scheduling and repair stages, and extend the framework to multi-language and larger-scale repositories through hierarchical SDS planning and cross-language interface contracts. We discuss these directions in detail in Section~\ref{subsec:disc-imp}.

\section*{Data Availability}
The replication package for this study is available at~\BlueNumber{\cite{replpack}}\DIFaddbegin \DIFadd{, including the run-level result tables, diagnostic summaries, and statistical analysis scripts used to reproduce all reported tables and statistical comparisons}\DIFaddend .

\begin{acks}
This work has been partially supported by the National Natural Science Foundation of China (NSFC) with Grant Nos. 92582203 and 62402348.%, and the Major Science and Technology Project of Hubei Province under Grant No. 2024BAA008.
\end{acks}

%\clearpage
\bibliographystyle{ACM-Reference-Format}
\bibliography{ref}

\end{document}